\documentclass[english]{article}
\usepackage[T1]{fontenc}
\usepackage[latin9]{inputenc}
\usepackage{geometry}
\usepackage{float}
\usepackage{amsmath}
\usepackage{amssymb}
\usepackage{graphicx}
\usepackage[authoryear]{natbib}
\usepackage{subfig}
\usepackage{mathrsfs}  \allowdisplaybreaks
\usepackage[symbol]{footmisc}
\usepackage{url}
\usepackage{xcolor}
\usepackage{bm}
\usepackage{mathrsfs}

\newcommand\indep{\protect\mathpalette{\protect\independenT}{\perp}}
\def\independenT#1#2{\mathrel{\rlap{$#1#2$}\mkern2mu{#1#2}}}

\newtheorem{theorem}{Theorem}\newtheorem{Assumption}{Assumption}\newtheorem{remark}{Remark}\newtheorem{definition}{Definition}\newtheorem{lemma}{Lemma}

\begin{document}


\begin{center}
\Large
\textbf{Outcome-Adjusted Balance Measure for Generalized Propensity Score Model Selection}
\vspace{1cm}

\large
Honghe Zhao\footnote[1]{Department of Statistics, North Carolina State University, 2311 Stinson Dr., Raleigh, NC 27695-8203, USA: hzhao22@ncsu.edu} and Shu Yang\footnote[2]{Department of Statistics, North Carolina State University, 2311 Stinson Dr., Raleigh, NC 27695-8203, USA: syang24@ncsu.edu}
\end{center}



\begin{abstract}
In this article, we propose the outcome-adjusted balance measure to perform model selection for the generalized propensity score (GPS), which serves as an essential component in estimation of the pairwise average treatment effects (ATEs) in observational studies with more than two treatment levels. The primary goal of the balance measure is to identify the GPS model specification such that the resulting ATE estimator is consistent and efficient. Following recent empirical and theoretical evidence, we establish that the optimal GPS model should only include covariates related to the outcomes. Given a collection of candidate GPS models, the outcome-adjusted balance measure imputes all baseline covariates by matching on each candidate model, and selects the model that minimizes a weighted sum of absolute mean differences between the imputed and original values of the covariates. The weights are defined to leverage the covariate-outcome relationship, so that GPS models without optimal variable selection are penalized. Under appropriate assumptions, we show that the outcome-adjusted balance measure consistently selects the optimal GPS model, so that the resulting GPS matching estimator is asymptotically normal and efficient. We compare its finite sample performance with existing measures in a simulation study. We illustrate an application of the proposed methodology in the analysis of the Tutoring data.
\end{abstract}


\section{Introduction}
\label{sec1}

The propensity score (PS) plays a central role in drawing causal conclusions from observational data. In their seminal work, \citet{rosenbaum1983} define PS as the conditional probability of receiving a treatment given baseline covariates, and show its effectiveness in removing confounding bias in estimating the average treatment effect (ATE). For settings with more than two categorical treatment levels, \citet{imbens2000} introduces the generalized propensity score (GPS) that naturally extends the definition of PS to facilitate estimation of the average potential outcome. \citet{yang2016} formalize the estimation and inference procedure for pairwise average treatment effects by proposing the GPS matching and subclassification estimators. Despite the popularity of these (G)PS based matching methods, one inevitable challenge lies in modeling and estimation of the (G)PS itself. 

Considerable progress has been made towards developing modeling strategies for the PS. The essential purpose a PS model specification serves is in assisting the estimation and inference of the ATE(s) rather than in explaning or predicting treatment assignment. Standard diagnostics for model prediction performance should be avoided as they often fail to suggest PS models that provide unbiased and efficient estimation of the ATE (\citealp{westreich2011}). This is supported by a large thread of empirical evidence, which surprisingly suggests that, including instrumental variables (covariates that are part of the true PS model) in the PS model specification inflates variance of the resulting ATE estimators (\citealp{brookhart2006,austin2007,myers2011,pearl2011}). Additionally, including precision variables (covariates only related to outcomes but not treatment) in the PS leads to improved efficiency for ATE estimation (\citealp{brookhart2006,patrick2011}). Building upon the work of \citet{lunceford2004} and \citet{hahn2004}, \citet{tang2020} theoretically justify that both excluding instrumental variables and including precision variables in the PS will help improve asymptotic efficiency in the Horvitz-Thompson estimator, the ratio estimator, and the doubly robust estimator. \citet{rotnitzky2019} provide a graphical criterion for identifying the optimal covariate adjustment set for non-parametric efficient estimation of the ATE.

When baseline covariates are high-dimensional, many regularized regression methods have been developed to perform variable selection for the PS model. \citet{shortreed2017} propose the outcome-adaptive LASSO with a penalty function that takes into account association between covariates and outcome, and association between covariates and treatment. \citet{ju2019} propose a collaborative-controlled LASSO that uses the C-TMLE algorithm based on LASSO to minimize a bias-variance tradeoff in the estimated treatment effect. \citet{tang2020} improve upon the outcome-adaptive LASSO by incorporating the ball covariance (\citealp{pan2020}), which makes the method free of dependence on the outcome model specification. These methods are useful not only for screening out redundant, highly correlated variables but also for arriving at a propensity score model that is consistent and efficient in ATE estimation. However, all of these methods technically must presuppose a parametric logit model for the PS, which could be restrictive given the primary objective is no longer to explain or predict treatment assignment. 

A reasonable diagnostic for the PS model is to assess the resulting covariate balance in the matched sample or balance within each stratum after stratifying on the quantiles of the PS. \citet{austin2007} investigate the issue of variable selection by comparing the ability of different PS model specifications in balancing baseline covariates. \citet{austin2009}, \citet{belitser2011}, and \citet{ali2014} carry out simulations that compare the ability of different balance measures (standardized mean difference, KS distance, etc.) in assessing whether a PS model is adequate in reducing finite sample bias.  However, most of these works (except Belitser and others) only focus on standard measures of balance, which assign equal weights to all covariates and hence do not make a distinction among types of covariates. \citet{caruana2015} propose a weighted standardized mean difference for PS variable selection, where the weights are coefficients from regressing the observed outcome on the covariates. 

We propose a recipe for GPS model selection for estimating pairwise ATEs in observational studies with multiple ($\geq2$) treatment levels. Motivated by the aforementioned literature, the optimal GPS model is the one that includes only covariates that are predictors of the potential outcomes. 
Targeting for the optimal GPS model, we propose the outcome-adjusted balance measure as a criterion for identifying the optimal model among a set of postulated models, given that set of models indeed contains the optimal model. The balance measure evaluates the discrepancies between two estimators of the average covariates, i.e., the GPSM estimator and the sample average of covariates, and imposes weights that penalize models with variable selection different from the optimal GPS model. 
The weights incorporate the association of covariates and outcome, which can be estimated either parametrically or nonparametrically. 
We show the selection consistency, i.e., the optimal model can be selected with probability one asymptotically, and 
that the resulting GPSM estimator of the ATEs based on this model selection criterion is not only consistent and asymptotically normal, but also efficient in estimation of the ATEs. 

The remaining article is organized as follows. We first review the observational studies with multiple treatment levels setup, along with definitions of the GPS and the GPSM estimator in Section \ref{sec2}. Section \ref{sec3} formally discusses details on balance assessment in the matched sample and construction of the outcome-adjusted balance measure. Section \ref{sec4} presents theoretical properties of the balance measure and the resulting GPSM estimator. We examine the finite sample estimation, inference, and model selection performance of the GPSM estimator by using the outcome-adjusted balance measure as a GPS model selection criterion in a Monte Carlo simulation study in Section \ref{sec5}. In Section \ref{sec6}, we implement the proposed balance measure in a real-world application. Finally, limitations of our method and possible future research directions are discussed in Section \ref{sec7}.

\section{Background}
\label{sec2}

\subsection{Setup}

Following the potential outcomes framework, consider an observational study with $T$ unordered treatment levels. Let $W\in\mathbb{W}=\{1,\ldots,T\}$ denote the treatment. For every unit in the population of interest, there are $T$ potential outcomes, denoted by $Y(w)$, for $w\in\mathbb{W}$, and \textit{$d$} observed baseline covariates $\boldsymbol{X}=\left(X^{(1)},...,X^{(d)}\right)^{\text{T}}$. Implicit in this setup is the stable-unit-treatment-value assumption, which states that there is no interference between units and no different versions of potential outcome for each treatment level. Define the indicator variable $D(w)\in\{0,1\}$ as $D(w)=
1$ if $W=w$, and $D(w)=0$ otherwise.

\begin{definition}[Generalized propensity score]\label{GPS}
The generalized
propensity score is the conditional probability of receiving each
treatment level: $p(w\mid\bm{x})=pr(W=w\mid\boldsymbol{X}=\boldsymbol{x}).$
\end{definition}

\begin{Assumption}[Overlap] \label{overlap}For all values of $\bm{x}$,
the probability of receiving any level of the treatment is positive: $p(w\mid\boldsymbol{x})>0$ for all $w,\boldsymbol{x}.$
\end{Assumption}

\begin{Assumption}[Weak unconfoundedness] \label{wu}For all $w\in\mathbb{W}$, $D(w)\ \indep\ Y(w)\ \mid\ \boldsymbol{X}.$
\end{Assumption}

Assumption \ref{overlap} rules out deterministic treatment assignment mechanisms and allows all units to have positive probabilities of receiving any treatment level.   
When Assumption \ref{overlap} is violated, it implies that there is a sub-population for which no information on some potential outcomes is available. Assumption \ref{wu} holds if all  baseline covariates that are associated with both treatment assignment and the outcome are captured. Therefore, in order to make Assumption \ref{wu} hold, practitioners often collect a rich set of pre-treatment variables, rendering variable selection a critical matter to consider. 

As a result of Assumptions \ref{overlap} and \ref{wu}, weak unconfoundedness is preserved if we condition on the generalized propensity score:
\begin{equation}
D(w)\ \indep\ Y(w)\ \mid\ p(w\mid\boldsymbol{X})\label{eq:WC_y}
\end{equation}

One key insight from \citet{yang2016} is that the conditional independence (\ref{eq:WC_y}) is sufficient for the identification of the average potential outcomes for a single treatment level, namely $\mathbb{E}\left\{ Y(w)\right\} $ for all $w$, in that $\mathbb{E}\left\{ Y(w)\right\}=\mathbb{E}[\mathbb{E}\left\{ Y\mid W=w, p(w\mid\boldsymbol{X}) \right\}].$ This in turn makes the identification of the pairwise average causal effects $\mathbb{E}\left\{Y(w')-Y(w)\right\}$ feasible. 

In the following discussion, suppose that we observe a random sample $\left\{ \boldsymbol{X}_{i},W_{i},Y_{i}(1),...,Y_{i}(T)\right\} $ from the population, where $i=1,...,N$, and suppose that all covariates have been normalized to have mean zero and variance one.

\subsection{Matching on the generalized propensity score}

We briefly review the generalized propensity score matching (GPSM) estimator of the pairwise average treatment effects.  Different from matching algorithms that construct matches only for individuals in a particular treatment group without replacement, here matching is done with replacement to impute the $T-1$ missing potential outcomes for every unit in the sample. Define the generalized propensity score matching function as $m(w,p)=\underset{j:W_{j}=w}{\text{argmin}}\:||p(w\mid\boldsymbol{X}_{j})-p||.$
Given the generalized propensity score matching function, we impute $Y_{i}(w)$ as $\widehat{Y}_{i}(w)=Y_{m\left\{ w,p(w\mid\boldsymbol{X}_{i})\right\} }.$
That is, for each treatment $w$, we impute the potential outcome of unit $i$ by the observed outcome from a unit in treatment $w$ that has generalized propensity score $p(w\mid\boldsymbol{X})$ most similar to that of unit $i$. We formulate the GPSM estimator of $\mathbb{E}\{Y(w)\}$ as $\widehat{\mathbb{E}}\left\{ Y(w)\right\} =N^{-1}\sum\limits _{i=1}^{N}\widehat{Y}_{i}(w).$
The final GPSM estimator of the pairwise average treatment effect is 
\begin{equation}
\widehat{\tau}_{{\rm gpsm}}(w,w')=\widehat{\mathbb{E}}\left\{ Y(w')\right\} -\widehat{\mathbb{E}}\left\{ Y(w)\right\} .\label{eq:gpsm}
\end{equation}

\citet{yang2016} show that including all confounders (covariates that are associated with both treatment assignment and potential outcome) in the GPS is sufficient for $\widehat{\tau}_{{\rm gpsm}}(w,w')$ to be asymptotically normal and consistent for the true pairwise ATE, i.e. $\mathbb{E}\left\{ Y(w')-Y(w)\right\}$.  In practice, to ensure the key unconfoundedness assumption holds, a rich set of pre-treatment covariates are collected and used to estimate the GPS. However, such an approach completely ignores the consequence for efficiency loss, especially if one includes instrumental variables, or fails to include precision variables in the GPS specification. Moreover, severe misspecification of the functional form of the optimal GPS (such as using the wrong link function, or failure to include higher-order and interaction terms) could easily lead to biased estimation of the ATEs.

\section{Model Selection}
\label{sec3}

In this section, we describe in detail the rationales underlying the construction of the outcome-adjusted balance measure. 

\subsection{Imputing covariates via GPS matching}

The GPSM estimator has been used to estimate the ATEs, but we propose to use it for estimating the mean of $\bm{X}$. This might appear bizarre and unnecessary, since $\mathbb {E}(\bm{X})$ can already be directly estimated by the sample average. The key insight is that the GPSM estimator of $\mathbb {E}(\bm{X})$ becomes useful for gauging the quality of a GPS model specification.  The following lemmas provide the stepping stones for constructing the proposed balance measure to select the optimal GPS model.

\begin{lemma}\label{Lemma1}Let $\boldsymbol{X}$ be any subset of measured baseline covariates, and let $X\in\boldsymbol{X}$ be any single covariate in this subset. Then for all $w\in\mathbb{W}$, we have
\begin{equation}
D(w)\ \indep\ X\ \mid\ p(w\mid\boldsymbol{X}).\label{eq:WC_x}
\end{equation}
\end{lemma}

Note that (\ref{eq:WC_x}) is parallel to (\ref{eq:WC_y}). Because of (\ref{eq:WC_y}), the imputed potential outcome for each treatment level is representative of the true potential outcome distribution over the whole sample for that treatment level. Similarly, because of (\ref{eq:WC_x}), the imputed covariate for treatment level $w$, namely, $\widehat{X}_{i}(w)=X_{m\{w,p(w\mid\boldsymbol{X}_{i})\}}$.
is representative of the covariate distribution over the whole sample for that treatment level. We define the GPSM estimator of $\mathbb{E}(X)$ as the sample average of imputed covariate for treatment level $w$, namely, $\widehat{X}_{{\rm gpsm}}(w)=N^{-1}\sum\limits _{i=1}^{N}\widehat{X}_{i}(w).$
Assuming the regularity conditions for applying the martingale central limit theorem hold (see Supplementary Material). Denote the sample average of covariate $X$ as $\overline{X}=N^{-1}\sum\limits_{i=1}^{N}X_{i}$. We have the following asymptotic normality result for $\widehat{X}_{{\rm gpsm}}(w)$.

\begin{lemma} 
\label{Lemma:Let-all-confounders}For any $w\in\mathbb{W}$, let $\boldsymbol{X}$ and $X$ be such that (\ref{eq:WC_x}) holds. Use the generalized propensity score $p(w\mid\boldsymbol{X})$ as the matching variable. Then we have $\ensuremath{\sqrt{N}\left\{\widehat{X}_{{\rm gpsm}}(w) -\overline{X}\right\}\stackrel{d}{\rightarrow}\mathcal{N}\left(0,\sigma_{X}^{2}\right)}$,
where $\sigma_{X}^{2}=\mathbb{E}\left[\mathbb{V}\left\{ X\mid p\left(w\mid\boldsymbol{X}\right)\right\} \left\{ \frac{3}{2}\frac{1}{p\left(w\mid\boldsymbol{X}\right)}-\frac{1}{2}p\left(w\mid\boldsymbol{X}\right)-1\right\} \right].$

\end{lemma}

Thus, in order to assess the quality of a proposed GPS specification $p(w\mid\boldsymbol{X})$, it is meaningful to impute the covariates by matching on $p(w\mid\boldsymbol{X})$, and compare how close their imputed distributions are to their original sample distributions. 
Many balance metrics have been studied in regards to their ability to prevent PS model misspecification. Most notable metrics include the absolute mean difference, Kolmogorov-Smirnov distance, mahalanobis distance and the weighted balance measure (\citealp{austin2009,belitser2011,ali2014,caruana2015}). In the binary treatment setting, they are simply defined to measure discrepancies between covariate distributions in the control and treatment arms. For comparison, we extend their definitions to the multi-level treatment settings to measure the discrepancies between the imputed and the original covariate distributions, and their definitions are included in the Supplementary Material.

\subsection{Types of covariates}
\begin{figure}[h!]
\begin{centering}
\includegraphics{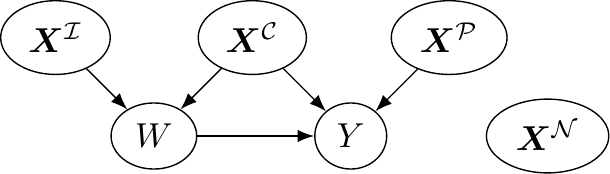}
\par\end{centering}
\caption{\label{fig:DAG}A DAG that illustrates the relationships among four types of covariates, treatment assignment, and outcome in a simple scenario.}
\end{figure}

For efficiency considerations, we have established that the optimal GPS model includes only outcome related pre-treatment variables. Therefore, in order to identify these variables, it is necessary to properly distinguish the types of covariates, and to examine the consequence of Lemma \ref{Lemma1} and Lemma \ref{Lemma:Let-all-confounders} in light of this distinction.
We consider a simple scenario where covariates can be categorized into four types, with causal relationships represented by the directed acyclic graph (Figure~\ref{fig:DAG}).

We let $\boldsymbol{X}^{\mathcal{C}}=\left\{ X^{(j)}\in pa(Y):W\text{ and }X^{(j)}\text{ are d-connected given }pa(Y)\backslash\left\{ X^{(j)}\right\} \right\} $ denote the confounders, where $pa(Y)$ denotes the parents of $Y$. We further define $\boldsymbol{X}^{\mathcal{P}}=pa(W)\backslash\boldsymbol{X}^{\mathcal{C}}$ as precision variables, $\boldsymbol{X}^{\mathcal{I}}=pa(W)\backslash\boldsymbol{X}^{\mathcal{C}}$ as instrumental variables, and $\boldsymbol{X}^{\mathcal{N}}=\boldsymbol{X}\backslash(\boldsymbol{X}^{\mathcal{C}}\cup\boldsymbol{X}^{\mathcal{P}}\cup\boldsymbol{X}^{\mathcal{N}})$ as noise variables. Refer to \citet{pearl2000} for a thorough discussion of the definitions of ``d-connected'' and ``parents'', as well as necessary assumptions involved in defining a DAG.

\begin{remark}If $X\in\boldsymbol{X}^{\mathcal{I}}\cup\boldsymbol{X}^{\mathcal{C}}$ is either an instrumental variable or a confounder, then (\ref{eq:WC_x}) holds true as long as $X$ is included in $p(w\mid\boldsymbol{X})$. When $X$ is not included in $p(w\mid\boldsymbol{X})$, (\ref{eq:WC_x}) does not hold because $X$ is a parent of $D(w)$. If $X\in\boldsymbol{X}^{\mathcal{P}}\cup\boldsymbol{X}^{\mathcal{N}}$ is either a precision variable or a noise variable, then (\ref{eq:WC_x}) holds regardless of whether $p(w\mid\boldsymbol{X})$ includes $X$ because the path from $X$ to $W$ is always blocked. 
\end{remark}

\begin{remark} 
(a) We highlight Lemma \ref{Lemma:Let-all-confounders} as the key result. Because the sample average $\overline{X}$ is consistent for the population mean $\mathbb{E}(X)$ independent of the GPS specification $p(w\mid\boldsymbol{X})$, we would expect the difference between the matching estimator $\widehat{X}_{{\rm gpsm}}(w) $ and the sample mean $\overline{X}$ to be small in large samples if (i) $X$ and $D(w)$ are independent conditional on $p(w\mid\boldsymbol{X})$ and (ii) $p(w\mid\boldsymbol{X})$ is correctly specified. The same logic would not apply to the potential outcomes. Since $Y(w)$ are missing for individuals who are not in treatment group $w$, it would be infeasible to gauge the difference between $\widehat{Y}(w)$ and $\bar{Y}(w)$ as sample size grows.

(b) Lemma \ref{Lemma:Let-all-confounders} relies on Lemma \ref{Lemma1}. Therefore if $X\in\boldsymbol{X}^{\mathcal{I}}\cup\boldsymbol{X}^{\mathcal{C}}$ is either an IV or a confounder, then $\widehat{X}_{{\rm gpsm}}(w) -\overline{X}\stackrel{p}{\rightarrow}0$ as long as $X$ is included in $p(w\mid\boldsymbol{X})$. If $X\in\boldsymbol{X}^{\mathcal{P}}\cup\boldsymbol{X}^{\mathcal{N}}$ is either a precision variable or a noise variable, then $\widehat{X}_{{\rm gpsm}}(w) -\overline{X}\stackrel{p}{\rightarrow}0$ regardless of whether $p(w\mid\boldsymbol{X})$ includes $X$.

\end{remark}
To further improve the efficiency of an ATE estimator, empirical literature suggests that the \textit{optimal} PS specification should include only confounders and precision variables (\citealp{brookhart2006,austin2007,myers2011,tang2020}). However, so far no theoretical results has justified which combination of the variables to include in the GPS model for the GPSM estimator to be efficient. Thus, we rely on existing empirical and theoretical findings for the PS model for other ATE estimators to establish that the optimal GPS specification should also include only the confounders and precision variables; i.e. $p(w\mid\bm{X}^{\mathcal{C}\cup \mathcal{P}})$. 

A model selection criterion based on minimizing the absolute mean difference will favor models that include $\boldsymbol{X}^{\mathcal{I}}\text{ and }\boldsymbol{X}^{\mathcal{C}}$ regardless of whether they include $\boldsymbol{X}^{\mathcal{P}}$ or $\boldsymbol{X}^{\mathcal{N}}$, since $\boldsymbol{X}^{\mathcal{P}}$ or $\boldsymbol{X}^{\mathcal{N}}$ will always be balanced as mentioned in Remark 2(b). As a result, the matching estimator based on a GPS model that minimizes the absolute mean difference would be consistent for the ATE but not necessarily efficient. Although $\mathtt{KSdist}$ is more sensitive to small differences in the shape of the distributions and $\mathtt{Mdist}$ takes into account the potential correlations among the covariates, these balance measures still place equal emphasis on balancing all baseline covariates, just like $\mathtt{AMD}$. As a result, they will likewise fail to encourage the inclusion of $\left\{ \boldsymbol{X}^{\mathcal{P}},\boldsymbol{X}^{\mathcal{C}}\right\} $ or discourage the inclusion of $\left\{ \boldsymbol{X}^{\mathcal{I}},\boldsymbol{X}^{\mathcal{N}}\right\} $ in a GPS model. While the weighted balance measure increases the chance of including prognostically important variables, it could still fail to exclude $\boldsymbol{X}^{\mathcal{I}}$ and $\boldsymbol{X}^{\mathcal{N}}$ in large samples due to the small weights they would receive. Nonetheless, as we will see next, it is possible to equip the absolute mean difference with outcome information to help identify the optimal GPS model.
\subsection{Outcome-adjusted balance measure}
To conform with evidence that advocates $p(w\mid\bm{X}^{\mathcal{C}\cup \mathcal{P}})$ as the optimal GPS
model, we make adjustments to the absolute mean difference
balance measure by leveraging the outcome information. 
We let $\rho\left(X^{(j)},Y\mid W=w\right)$
denote a generic metric of correlation between the observed outcome
$Y$ and $j\text{th}$ observed covariate $X^{(j)}$ conditional
on treatment level $W=w$. We let $\rho_{N}\left(X^{(j)},Y\mid W=w\right)$
denote its empirical version. 
We define the outcome-adjusted balance measure for the GPS model $p(w\mid\boldsymbol{X})$
at treatment level $w\in\mathbb{W}$ to be:
\begin{align*}
\mathtt{OABM}_\rho\left\{ w,p(w\mid\boldsymbol{X})\right\}  & =\left.\boldsymbol{\zeta}_{\rho}\left\{ w,p(w\mid\boldsymbol{X})\right\} \right.^{\mathrm{T}}\left|\widehat{\boldsymbol{X}}_{{\rm gpsm}}(w) -\overline{\boldsymbol{X}}\right|\\
 & =\sum\limits _{j=1}^{d}\zeta_{\rho}^{(j)}\left\{ w,p(w\mid\boldsymbol{X})\right\} \left|\widehat{X}^{(j)}_{{\rm gpsm}}(w) -\overline{X^{(j)}}\right|,
\end{align*}
with weights $\boldsymbol{\zeta}_{\rho}\left\{ w,p(w\mid\boldsymbol{X})\right\} =\left[\zeta_{\rho}^{(1)}\left\{ w,p(w\mid\boldsymbol{X})\right\} ,....,\zeta_{\rho}^{(d)}\left\{ w,p(w\mid\boldsymbol{X})\right\} \right]^{\text{T}}$
defined as follows: 

\[
\zeta_{\rho}^{(j)}\left\{ w,p(w\mid\boldsymbol{X})\right\} =\begin{cases}
1/\rho_{N}\left(X^{(j)},Y\mid W=w\right) & \text{if }X^{(j)}\in p(w\mid\boldsymbol{X})\\
\delta_{w}\rho_{N}\left(X^{(j)},Y\mid W=w\right) & \text{if }X^{(j)}\notin p(w\mid\boldsymbol{X})
\end{cases}.
\]
That is, if $X^{(j)}$ is included in $p(w\mid\boldsymbol{X})$, then
$\zeta_{\rho}^{(j)}\left\{ w,p(w\mid\boldsymbol{X})\right\} $ takes the value
$1/\rho_{N}(X^{(j)},Y\mid W=w)$; if $X^{(j)}$ is excluded
from the model $p(w\mid\boldsymbol{X})$, then $\zeta_{\rho}^{(j)}\left\{ w,p(w\mid\boldsymbol{X})\right\}$ assumes the value $\delta_{w}\rho_{N}(X^{(j)},Y\mid W=w)$. We let $\delta_{w}$ be a positive tuning parameter that is proportional to $N^{1/3}$. 

The purpose of the weighting design $\boldsymbol{\zeta}_{\rho}\left\{ w,p(w\mid\boldsymbol{X})\right\} $
is to penalize posited models that differ in variable selection from
the optimal GPS model, which should only include $\boldsymbol{X}^{\mathcal{P}}$
and $\boldsymbol{X}^{\mathcal{C}}$. Consider a strong predictor $X^{(j)}$
of the outcome $Y$ at treatment level $w$, which would be reflected
by a large $\rho_{N}\left(X^{(j)},Y\mid W=w\right)$. If $X^{(j)}$
is excluded from a candidate GPS model and therefore unbalanced,
we impose a large penalty $\delta_{w}\rho_{N}\left(X^{(j)},Y\mid W=w\right)$.
On the contrary, if we learn that $\rho_{N}\left(X^{(j)},Y\mid W=w\right)$
is small, this will indicate that $X^{(j)}$ is weakly correlated
or uncorrelated with the outcome. In that case, if $X^{(j)}$ is included
in a candidate GPS model and therefore balanced, we penalize such
a model by imposing a large weight $1/\rho_{N}\left(X^{(j)},Y\mid W=w\right)$. 

The choice of the correlation metric $\rho$ should depend on
whether one is confident in correctly specifying the parametric outcome
model. For instance, if one has strong knowledge that the potential
outcome $Y(w)$ is linear in the covariates with partial regression
coefficients $(\theta_w^{(1)},...,\theta_w^{(d)})$, one should let $\rho_{N}\left(X^{(j)},Y\mid W=w\right)=\left|\widehat{\theta}_{w}^{(j)}\right|$ and select the GPS model that minimizes $\mathtt{OABM}_{\mathtt{OLS}}$.
On the contrary, if one knows little about the relationship between
$Y(w)$ and $\boldsymbol{X}$, it is recommended to use the ball correlation (\citealp{pan2020}) as the correlation metric given that it is free of dependence on the modeling assumptions, and one should choose the GPS model that minimizes $\mathtt{OABM}_{\mathtt{BCor}}$. Other model-free correlation metrics, such as the distance correlation (\citealp{szekely2007}), are also feasible alternatives for $\rho$.

The role of the tuning parameter $\delta_{w}$ is to ensure adequate
finite sample performance. 
If one is confident about characterizing
the outcome model, choosing $\delta_{w}=N^{1/3}$ would be sufficient. If
this is not the case, and one uses the ball correlation as the metric
of correlation, then for every treatment level $w$, we first let  
$BCor_N^* =\{\max_{j:X^{(j)}\in\bm{X}^{\mathcal{H}_{0}}}BCor_{N}(X^{(j)},Y\mid W=w)+\min_{j:X^{(j)}\in\bm{X}^{\mathcal{H}_{1}}}BCor_{N}(X^{(j)},Y\mid W=w)\}/2,$
and then choose $\delta_{w}$ by minimizing $\left|1/BCor_N^*-\delta_{w}BCor_N^*\right|.$
Here, $\bm{X}^{\mathcal{H}_{0}}$ consists of baseline covariates
for which the ball covariance test between each covariate $X^{(j)}$
and $Y$ conditional on $W=w$ fails to reject the null hypothesis,
and $\bm{X}^{\mathcal{H}_{1}}$ is defined as the baseline covariates
for which the ball covariance test between each covariate $X^{(j)}$
and $Y$ conditional on $W=w$ rejects the null hypothesis. Therefore,
$\max_{j:X^{(j)}\in\bm{X}^{\mathcal{H}_{0}}}BCor_{N}(X^{(j)},Y\mid W=w)$
corresponds to the largest empirical ball correlation value among
variables tested to be unrelated to $Y$ at treatment level $w$,
whereas $\min_{j:X^{(j)}\in\bm{X}^{\mathcal{H}_{1}}}BCor_{N}(X^{(j)},Y\mid W=w)$
refers to the smallest empirical ball correlation value among variables
tested to be related to $Y$ at treatment level $w$, and $BCor_N^*$ is therefore a threshold ball correlation value that separates the covariates that are truly related to the outcome from those that are not.

The outcome-adjusted balance measure is computed separately for each treatment level $w$.
Therefore it is possible that for a different treatment arm, relevant baseline covariates will be different, and a different
GPS model will minimize $\mathtt{OABM}_{\rho}$ corresponding to that
arm. The selected GPS model for treatment level $w$ can then be used
to construct a point estimate for the mean potential outcome $\mathbb{E}\left\{ Y(w)\right\} $,
and then for the pairwise treatment effects. For variance estimation,
we recommend using the \citet{abadie2016} variance
estimator, with slight adjustments made to allow the possibility that
different GPS models could be selected by $\mathtt{OABM}_{\rho}$ for
different treatment levels. The formula for the variance estimator is contained in the Supplementary Material. We also summarize the steps we take to perform GPS model selection in the Supplementary Material.

\section{Theoretical Properties}
\label{sec4}

In this section, we present asymptotic results for the outcome-adjusted balance measure
as well as for the resulting GPSM estimator. Define the conditional
mean and variance of $Y(w)$ given $p(w\mid\bm{X})$ as $\bar{\mu}(w,p)=\mathbb{E}\left\{ Y(w)\mid p(w\mid\bm{X})=p\right\} $
and $\bar{\sigma}^{2}(w,p)=\mathbb{V}\left\{ Y(w)\mid p(w\mid\bm{X})=p\right\} $.
Define the mean potential outcome as $\mu(w)=\mathbb{E}\left\{ Y(w)\right\} $.
Let $p(w\mid\boldsymbol{X}^{\mathcal{C}\cup\mathcal{P}})$ denote
the optimal generalized propensity score model. 

In addition to the assumptions made in Section \ref{sec2},
model selection consistency of $\mathtt{OABM}_{\rho}$ also relies on
correct specification of the parametric outcome model.

\begin{Assumption}[Correct specification of parametric outcome model] 
\label{OM}

For all $w\in\mathbb{W}$, the relationship between the potential
outcome $Y_{i}(w)$ and covariates $\bm{X}_{i}$ can be characterized
by a known parametric density/mass function $f_{w}(y_{i}\mid\bm{x}_{i},\bm{\theta},\bm{\phi})$,
$i=1,...,N$, where the parameters of interest $\bm{\theta}\in\Theta\subset\mathbb{R}^{d}$ are associated with the effect of $\bm{X}_i$, and $\bm{\phi}$ are the nuisance parameters. 

\end{Assumption} 
\begin{theorem}[Model selection consistency] \label{theorem1-model selection consistency}

Let the collection of posited generalized propensity score models
be $\mathscr{P}$ and suppose the optimal model $p(w\mid\bm{X}^{\mathcal{C}\cup \mathcal{P}})\in\mathscr{P}$
is one of the posited models. Suppose Assumptions \ref{overlap} and \ref{wu} hold. 

(i) Suppose also that Assumption \ref{OM} is satisfied. Let $\rho_{N}(X^{(j)},Y\mid W=w)$
be the absolute value of a CAN (consistent and asymptotic normal) estimator of the true effect of covariate $X^{(j)}$ on $Y(w)$. 
Then for all $w$, we have $\lim_{n\rightarrow\infty}\mathbb{P}[ \mathtt{OABM}_{\mathtt{CAN}}\left\{w,p(w\mid\bm{X}^{\mathcal{C}\cup \mathcal{P}})\right\}\leq\mathtt{OABM}_{\mathtt{CAN}}(w,p_{k})\text{ for }p_{k}\in\mathcal{\mathscr{P}}] =1.$

(ii) Suppose that $\rho(X^{(j)},Y\mid W=w)=o(1)$. Let $\rho_{N}(X^{(j)},Y\mid W=w)$ be the empirical ball correlation between $X^{(j)}$ on $Y$ conditional on $W=w$. Then for all $w$, we have $\lim_{n\rightarrow\infty}\mathbb{P}[ \mathtt{OABM}_{\mathtt{BCor}}\left\{w,p(w\mid\bm{X}^{\mathcal{C}\cup \mathcal{P}})\right\}\leq\mathtt{OABM}_{\mathtt{BCor}}(w,p_{k})\text{ for }p_{k}\in\mathcal{\mathscr{P}}] =1.$
\end{theorem}
\begin{remark}
When $\rho_{N}$ is a nonparametric correlation metric such as the
ball correlation or the distance correlation, the above consistency
result holds approximately because $W$ is a collider of $\bm{X}^{\mathcal{I}}$ and $\bm{X}^{\mathcal{C}}$ (see Figure \ref{fig:DAG}). A consequence of this is that in general $\rho(X^{(j)},Y\mid W=w)\neq0$ for $X^{(j)}\in\boldsymbol{X}^{\mathcal{I}}$. In such cases, the
reliability of $\mathtt{OABM_{BCor}}$ depends on how small such correlation
$\rho(X^{(j)},Y\mid W=w)$ is in reality. 
\end{remark}
Theorem \ref{theorem1-model selection consistency} is the main result, which states that if one is capable
of correctly specifying the outcome model or collider bias is negligible (i.e. $\rho(X^{(j)},Y\mid W=w)=o(1)$), then the outcome-adjusted balance
measure is guaranteed to identify the optimal GPS model among a set
of posited models in large samples.

\citet{yang2016} prove that when the true
GPS has a multinomial logit form and is estimated using maximum likelihood,
the GPSM estimator matching on the estimated GPS is consistent and
asymptotically normal. We combine this result with Theorem \ref{theorem1-model selection consistency}
and conclude in Theorem \ref{theorem2} below that matching on the optimal GPS model selected by the balance measure results in a GPSM estimator that is consistent, asymptotically
normal and efficient for the average potential outcomes. 

Consider the following generic parametric form of the selected optimal GPS model. Let treatment level $1$ be the reference (baseline) category. Suppose a function of all other category relative to the baseline the follows a generalized linear form, i.e. $p(w\mid\bm{X}^{\mathcal{C}\cup\mathcal{P}}=\bm{x};\bm{\beta})=p\left(w\mid\bm{x}^{\text{T}}\bm{\beta}_{2},\ldots,\bm{x}^{\text{T}}\bm{\beta}_{T}\right)$ for all $w$. 
Let $I_{\bm{\beta}}$ be the information matrix. We estimate $\bm{\beta}$
using maximum likelihood, and denote the estimated GPS as $p(w\mid\boldsymbol{X}^{\mathcal{C}\cup\mathcal{P}};\widehat{\bm{\beta}})$.
Define the GPSM estimator of $\mu(w)$ that matches on $p(w\mid\boldsymbol{X}^{\mathcal{C}\cup\mathcal{P}};\widehat{\bm{\beta}})$
as $N^{-1}\sum_{i=1}^{N}Y_{m\left\{ w,p(w\mid\boldsymbol{X}_{i}^{\mathcal{C}\cup\mathcal{P}};\widehat{\bm{\beta}})\right\} }.$

For example, assume the parametric model is
a multinomial logit model.
 That is, for $w=\{2,...,T\}$, we assume the following model: $p(w\mid\bm{x};\bm{\beta})=p\left(w\mid\bm{x}^{\text{T}}\bm{\beta}_{2},\ldots,\bm{x}^{\text{T}}\bm{\beta}_{T}\right)=\exp\left(\bm{x}^{\text{T}}\bm{\beta}_{w}\right)\times\left\{ 1+\sum_{w^{\prime}=2}^{T}\exp\left(\bm{x}^{\text{T}}\bm{\beta}_{w'}\right)\right\} ^{-1}$, where $\bm{\beta}^{\text{T}}=(\bm{\beta}^{\text{T}}_{2},...,\bm{\beta}^{\text{T}}_{T})$ and $p(1\mid\bm{x};\bm{\beta})=\left\{ 1+\sum_{w^{\prime}=2}^{T}\exp\left(\bm{x}^{\text{T}}\bm{\beta}_{w'}\right)\right\} ^{-1}.$

\begin{Assumption} \label{assumption4} The optimal GPS model $p(w\mid\boldsymbol{X}^{\mathcal{C}\cup\mathcal{P}};\bm{\beta})$
has a continuous distribution with compact support $[\underline{p},\overline{p}]$
and with a continuous density function. The conditional expectation
of potential outcome $\bar{\mu}(w,p)$ is Lipschitz-continuous in
$p$. For some $\delta>0$, $\mathbb{E}\{ |Y|^{2+\delta}|W=w,p(w\mid\boldsymbol{X}^{\mathcal{C}\cup\mathcal{P}};\bm{\beta})=p\} $
is uniformly bounded.

\end{Assumption}
\begin{theorem}[Asymptotic normality of GPSM estimator based on optimal GPS model] \label{theorem2} Suppose all
assumptions made in Theorem \ref{theorem1-model selection consistency} are satisfied. Suppose also that Assumption \ref{assumption4} is satisfied. Then $p(w\mid\boldsymbol{X}^{\mathcal{C}\cup\mathcal{P}};\bm{\beta})$
is the GPS model selected by minimizing the outcome-adjusted balance measure at treatment
level $w$, and 
\[
N^{1/2}\left[N^{-1}\sum_{i=1}^{N}Y_{m\left\{ w,p(w\mid\boldsymbol{X}_{i}^{\mathcal{C}\cup\mathcal{P}};\widehat{\bm{\beta}})\right\} }-\mu(w)\right]\stackrel{d}{\rightarrow}\mathcal{N}\left\{ 0,\sigma^{2}(w)-\bm{c}\left(w\right)^{\text{T}}I_{\bm{\beta}}^{-1}\bm{c}\left(w\right)\right\} 
\]
where 
\begin{align*}
\sigma^{2}(w)= & \mathbb{E}\left[\bar{\sigma}^{2}\left\{ w,p(w\mid\boldsymbol{X}^{\mathcal{C}\cup\mathcal{P}};\bm{\beta})\right\} \left\{ \frac{3}{2}\frac{1}{p(w\mid\boldsymbol{X}^{\mathcal{C}\cup\mathcal{P}};\bm{\beta})}-\frac{1}{2}p(w\mid\boldsymbol{X}^{\mathcal{C}\cup\mathcal{P}};\bm{\beta})\right\} \right]\\
 & +\mathbb{E}\left(\left[\bar{\mu}\left\{ w,p(w\mid\boldsymbol{X}^{\mathcal{C}\cup\mathcal{P}};\bm{\beta})\right\} -\mu(w)\right]^{2}\right).
\end{align*}

\end{theorem}
The first term $\sigma^{2}(w)$ is the asymptotic variance associated
with matching on the true optimal generalized propensity score, and
$\bm{c}\left(w\right)^{\text{T}}I_{\bm{\beta}}^{-1}\bm{c}\left(w\right)$
corresponds to the gain in precision when matching variable is the
estimated optimal GPS. We rely on empirical evidence to conclude that
$\sigma^{2}(w)-\bm{c}\left(w\right)^{\text{T}}I_{\bm{\beta}}^{-1}\bm{c}\left(w\right)$
is smaller than the asymptotic variance corresponding to any other
GPS model specifications. The precise definition of $\bm{c}\left(w\right)$ can be found in the Supplementary Material.

\section{A Simulation Study}
\label{sec5}

In this section, we examine the finite sample performance of the outcome-adjusted balance
measure in a simulation study with three treatment levels. For each
simulated dataset, we generate nine independent standard normal covariates
$\bm{X}_{i}=\left(X_{i}^{(1)},...,X_{i}^{(9)}\right)^{\text{T}}$
for each individual $i$. The type of each covariate is specified
as follows: $\boldsymbol{X}^{\mathcal{C}}=(X^{(1)},X^{(2)})^{\text{T}}$,
$\boldsymbol{X}^{\mathcal{I}}=(X^{(3)},X^{(4)},X^{(5)})^{\text{T}}$,
$\boldsymbol{X}^{\mathcal{P}}=(X^{(6)},X^{(7)},X^{(8)})^{\text{T}}$,
$\boldsymbol{X}^{\mathcal{N}}=(X^{(9)})^{\text{T}}$. The treatment
indicators $\left\{D_{i}(1),D_{i}(2),D_{i}(3)\right\}$ are generated
from a multinomial distribution with event probabilities 
\[
p(W_{i}=w|\boldsymbol{X}_{i})=\exp\left\{ \left(1,\boldsymbol{X}_{i}\right)^{\text{T}}\bm{\beta}_{w}\right\} /\sum_{w^{\prime}=1}^{3}\exp\left\{ \left(1,\boldsymbol{X}_{i}\right)^{\text{T}}\bm{\beta}_{w^{\prime}}\right\} 
\]
for $w=\{1,2,3\}$. Here, $D_{i}(w)=1$ if unit $i$ belongs to treatment
$w$. We set $\bm{\beta}_{1}^{T}=(0,0,0,0,0,0,0,0,0,0)$, $\bm{\beta}_{2}^{T}=0.7\times(0,1.5,1.5,u,u,u,0,0,0,0)$,
and $\bm{\beta}_{3}^{T}=0.4\times(0,1.5,1.5,u,u,u,0,0,0,0)$, where
$u=1,2$ controls the strength of instrumental variables. The sample
size for each treatment is $N_{w}=500$, for $w=\{1,2,3\}$.

We consider two common ways potential outcomes may be related to the
covariates. First, we let the mean potential outcome
be a linear function of the covariates so that $Y_{i}(w)=\left(1,\boldsymbol{X}_{i}\right)^{\text{T}}\bm{\theta}_{w}+\eta_{i}$
with $\eta_{i}\sim \mathcal{N}(0,1)$. Second, we consider potential outcomes
generated from a nonlinear function of the covariates, that is $Y_{i}(w)=\left\{ 1,\left(\boldsymbol{X}_{i}^{2}-1\right)/2\right\} ^{\text{T}}\bm{\theta}_{w}+\eta_{i}$
with $\eta_{i}\sim \mathcal{N}(0,1)$. We let $\bm{\theta}_{1}^{T}=(-1.5,1.5,1.5,0,0,0,v,v,v,0)$,
$\bm{\theta}_{2}^{T}=(-3,1.5,1.5,0,0,0,v,v,v,0)$, and $\bm{\theta}_{3}^{T}=(1.5,1.5,1.5,0,0,0,-v,-v,-v,0)$,
where $v=1,2$ controls the strength of relationship between precision variables and potential outcomes. The true
pairwise average treatment effects are therefore $\tau(1,2)=-1.5,\tau(1,3)=3,\tau(2,3)=4.5$. 

We propose six GPS models that include the same covariates across
1,000 simulated datasets. Model $\mathtt{p(\bm{X}^{C})}$ includes only the confounders
$\boldsymbol{X}^{\mathcal{C}}$. Model $\mathtt{p(\bm{X}^{IC})}$ includes both confounders
$\boldsymbol{X}^{\mathcal{C}}$ and instrumental variables $\boldsymbol{X}^{\mathcal{I}}$.
Model $\mathtt{p(\bm{X}^{CP})}$ includes confounders as well as covariates that
are only related to outcome (i.e., $\boldsymbol{X}^{\mathcal{C}}$,
$\boldsymbol{X}^{\mathcal{P}}$). Model $\mathtt{p(\bm{X}^{all})}$ includes all nine
covariates. Model $\mathtt{p(\bm{X}^{1357})}$ includes covariates $(X^{(1)},X^{(3)},X^{(5)},X^{(7)})$, and model $\mathtt{p(\bm{X}^{2468})}$ includes $(X^{(2)},X^{(4)},X^{(6)},X^{(8)})$.
These six GPS models serve as benchmark models for comparing the performance
of $\mathtt{OABM}_{\rho}$ to other measures.

For each of the six GPS models with fixed covariates, the generalized
propensity scores are estimated using a multinomial logistic regression
model with covariates entering the model linearly. We carry out the
model selection procedure with two versions of $\mathtt{OABM}_{\rho}$ and
four other existing balance measures defined in the Supplementary Material:
$\mathtt{WBM}$, $\mathtt{AMD}$, $\mathtt{Mdist}$, and $\mathtt{KSdist}$.
$\mathtt{OABM}_{\mathtt{BCor}}$ is a variant of the outcome-adjusted balance measure with $\rho$
set to be the ball correlation. $\mathtt{OABM}_{\mathtt{OLS}}$ is another variant
$\mathtt{OABM}_{\rho}$ with $\rho$ chosen as the maximum likelihood estimator of the partial
regression coefficient from assuming that the potential outcomes are
linear in the covariates. Due to space constraints, here we only present results
for 4 of the 8 total scenarios. The remaining scenarios of the simulation
can be found in the Supplementary Material. 

\begin{figure}[h!]
\centering
\subfloat[\label{s2t2 linear}$u=2,v=1$\protect \\
Outcome \textit{linear} in covariates]{%
\includegraphics[width=7cm,height=7cm]{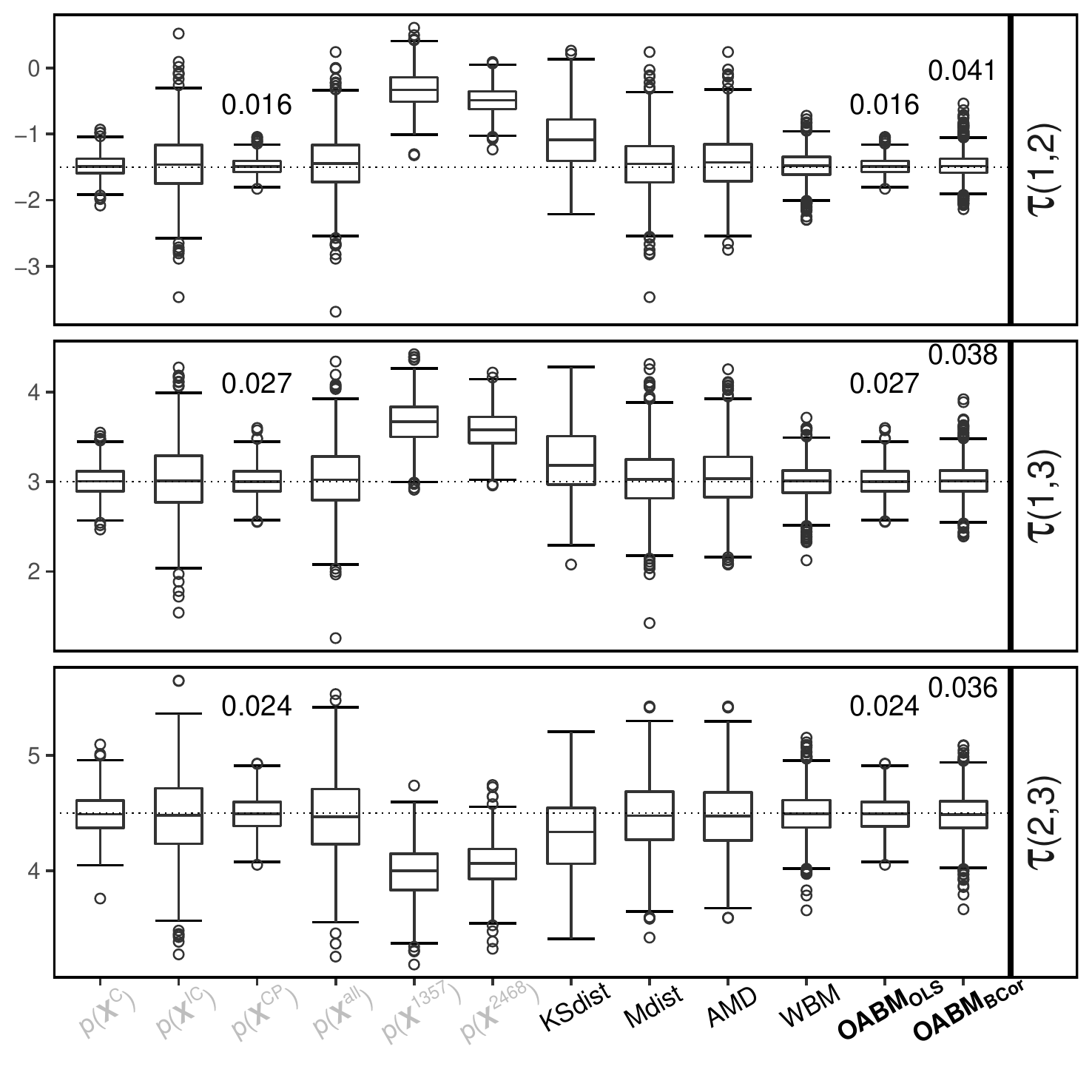}%
}%
\subfloat[\label{s2t3 linear}$u=2,v=2$\protect \\
Outcome \textit{linear} in covariates]{%
\includegraphics[width=7cm,height=7cm]{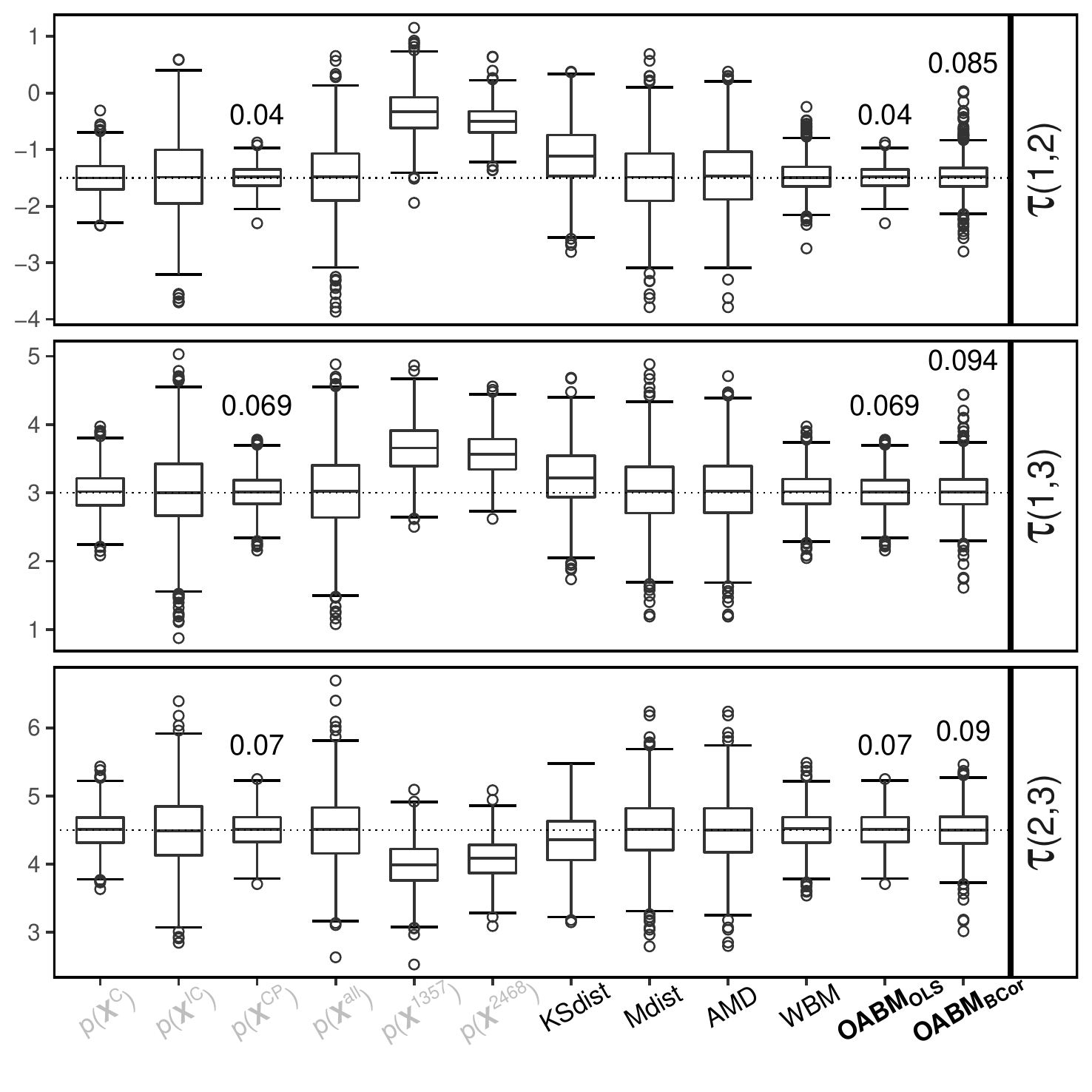}%
}%

\subfloat[$u=2,v=1$\protect \\
Outcome \textit{nonlinear} in covariates]{%
\includegraphics[width=7cm,height=7cm]{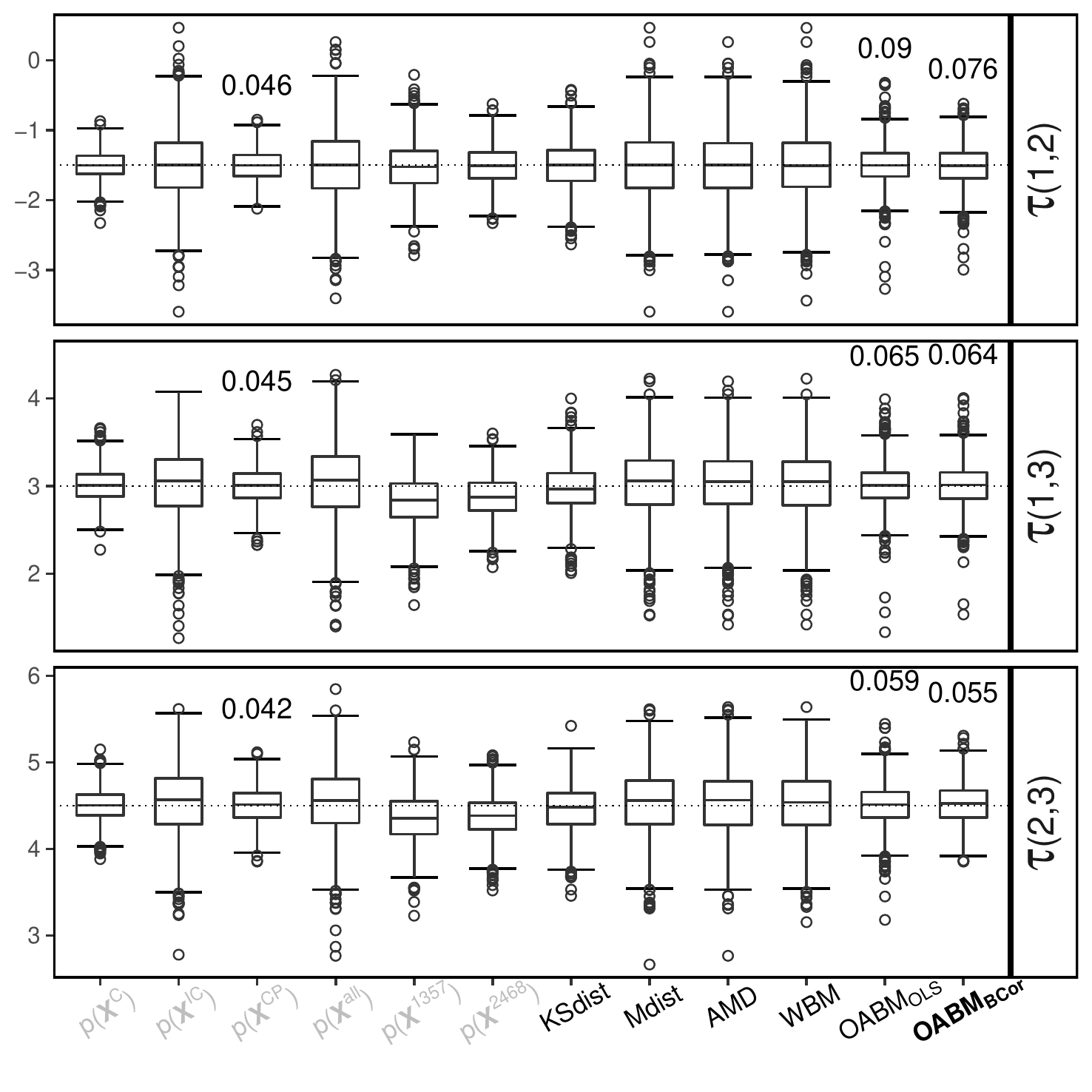}%
}%
\subfloat[\label{s2t3 nonlinear}$u=2,v=2$\protect \\
Outcome \textit{nonlinear} in covariates]{%
\includegraphics[width=7cm,height=7cm]{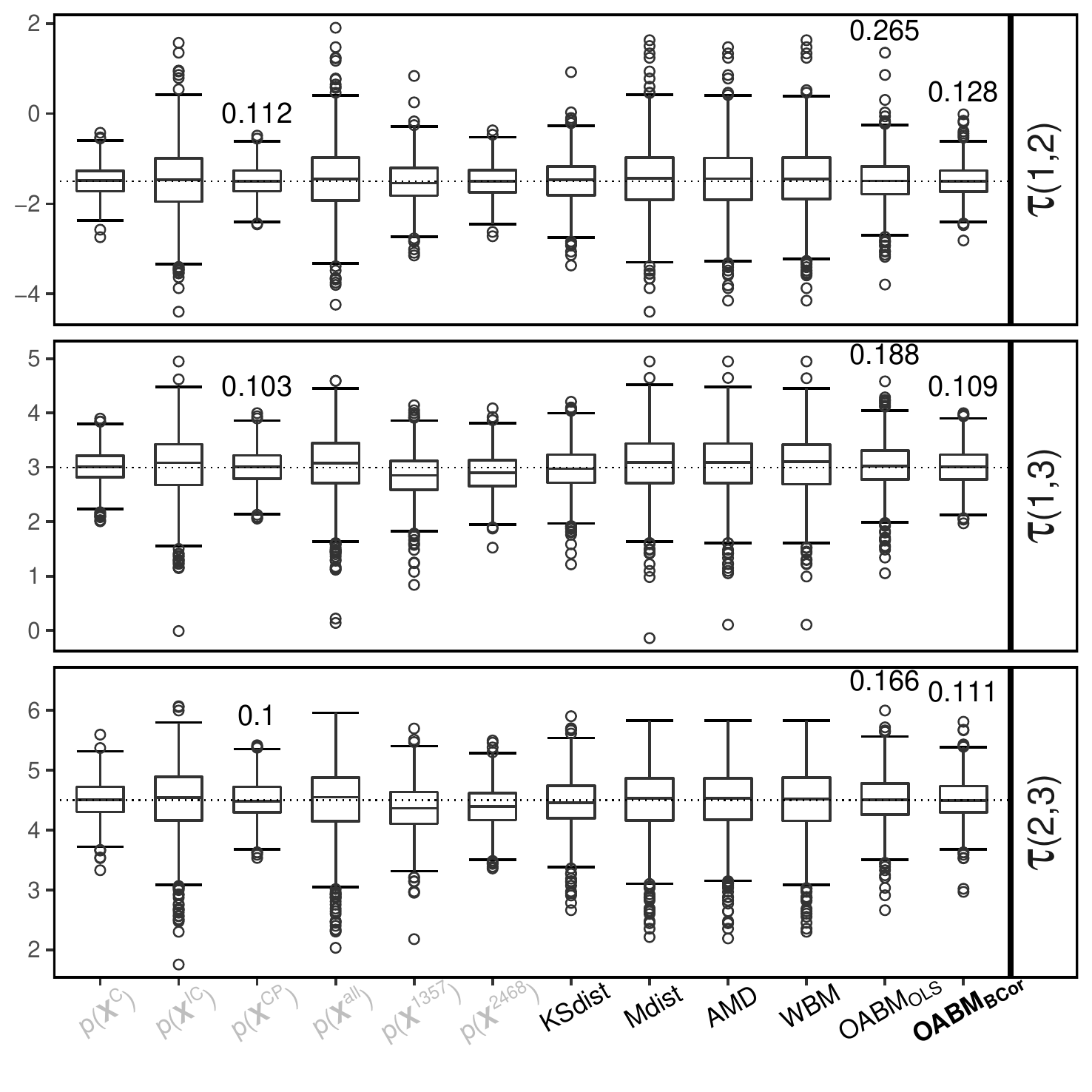}%
}%
\caption{\label{fig:boxplots}Box plots of 1000 generalized propensity score
matching estimates under 2 linear and 2 nonlinear scenarios. The
6 benchmark GPS models are greyed out on the $x$-axis. Numeric MSEs
for $\mathtt{p(\bm{X}^{CP})}$, $\mathtt{OABM}_{\mathtt{BCor}}$ and  $\mathtt{OABM}_{\mathtt{OLS}}$ are explicitly shown
above their corresponding box plots. True pairwise treatment effects
are denoted by the horizontal dotted lines.}
\end{figure}

\begin{figure}[h!]
\begin{centering}
\subfloat[$u=2,v=1$\protect \\
Outcome \textit{linear} in covariates]{\begin{centering}
\includegraphics[width=7cm,height=7cm]{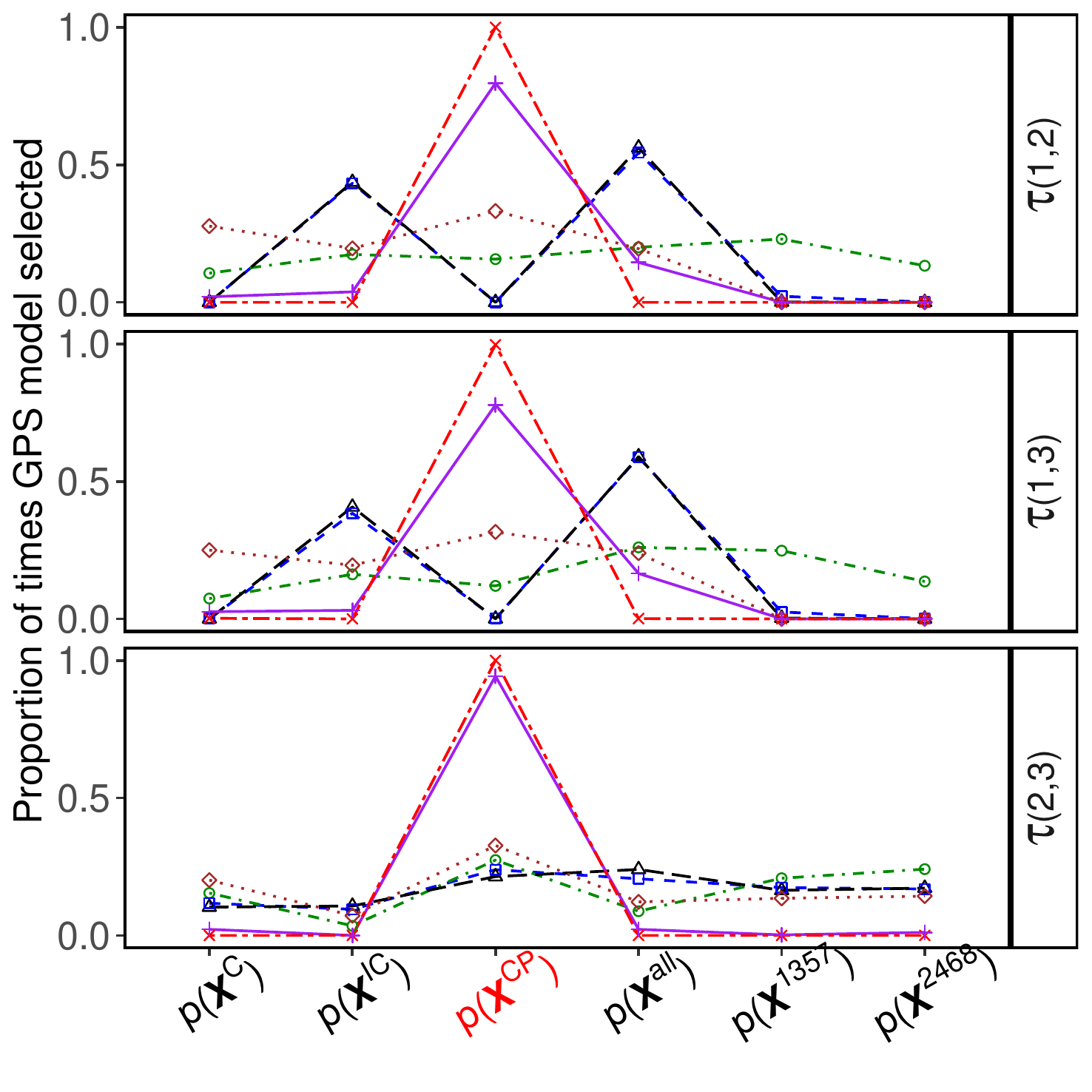}
\par\end{centering}
}\subfloat[$u=2,v=2$\protect \\
Outcome \textit{linear} in covariates]{\begin{centering}
\includegraphics[width=7cm,height=7cm]{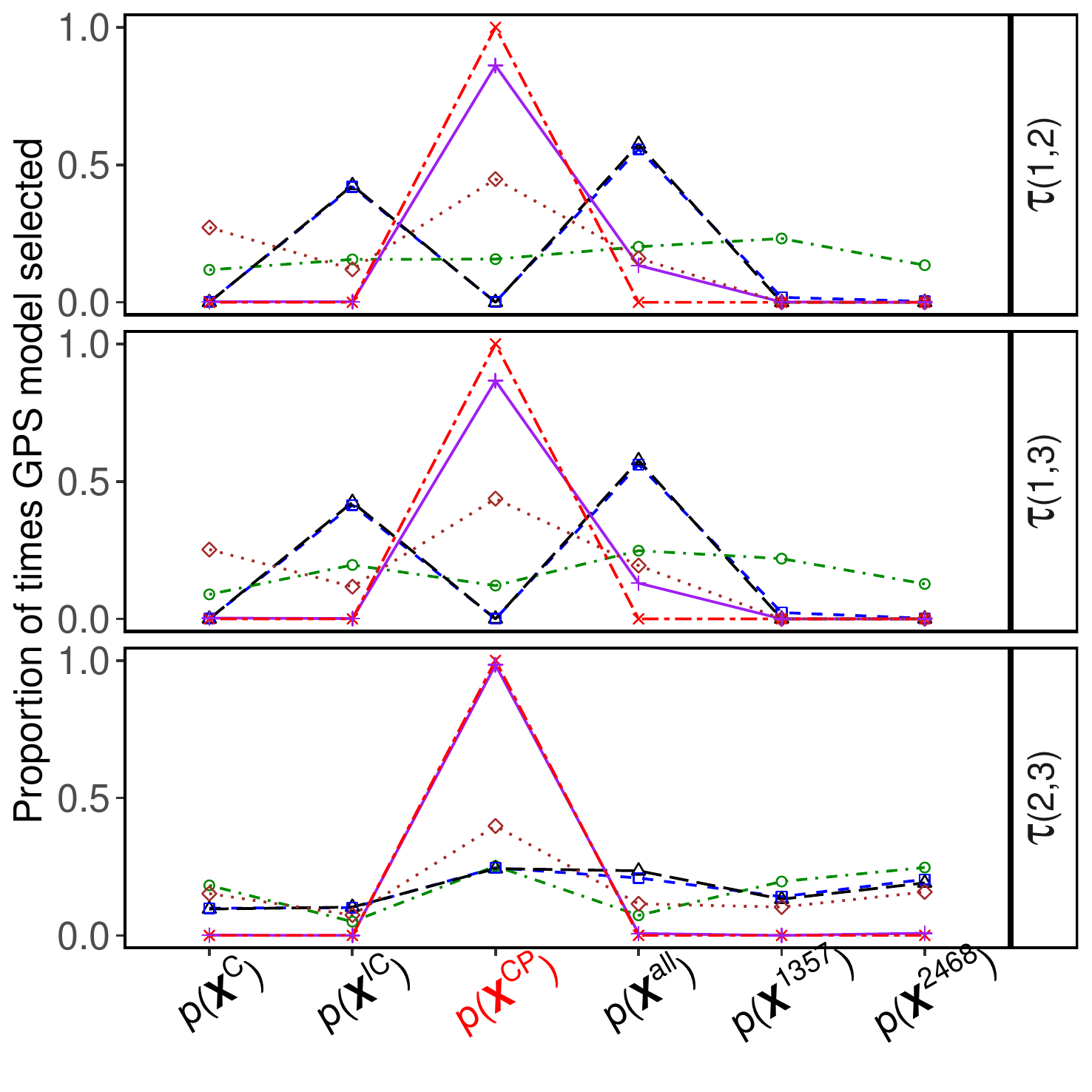}
\par\end{centering}
}
\par\end{centering}
\begin{centering}
\subfloat[$u=2,v=1$\protect \\
Outcome \textit{nonlinear} in covariates]{\begin{centering}
\includegraphics[width=7cm,height=7cm]{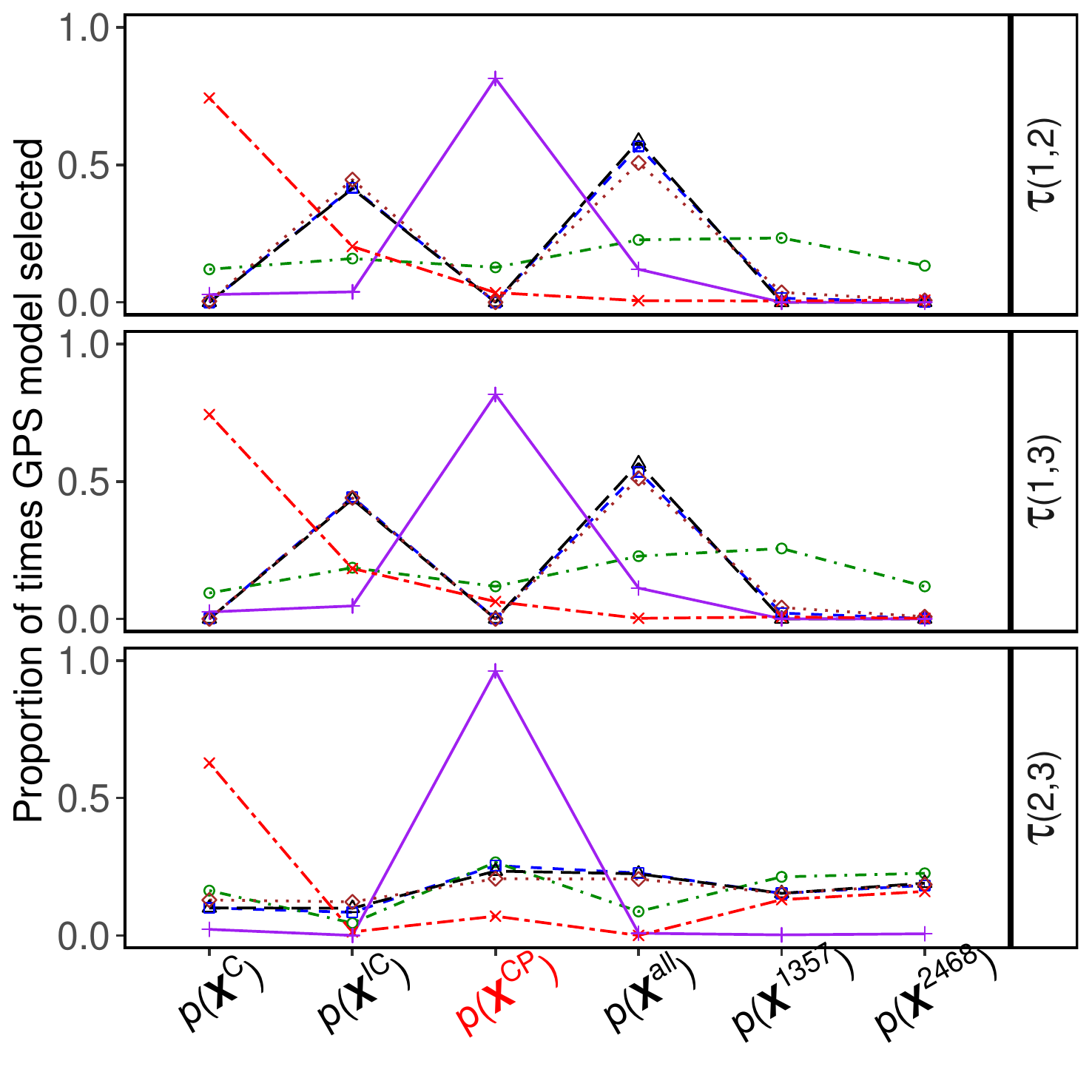}
\par\end{centering}
}\subfloat[$u=2,v=2$\protect \\
Outcome \textit{nonlinear} in covariates]{\begin{centering}
\includegraphics[width=7cm,height=7cm]{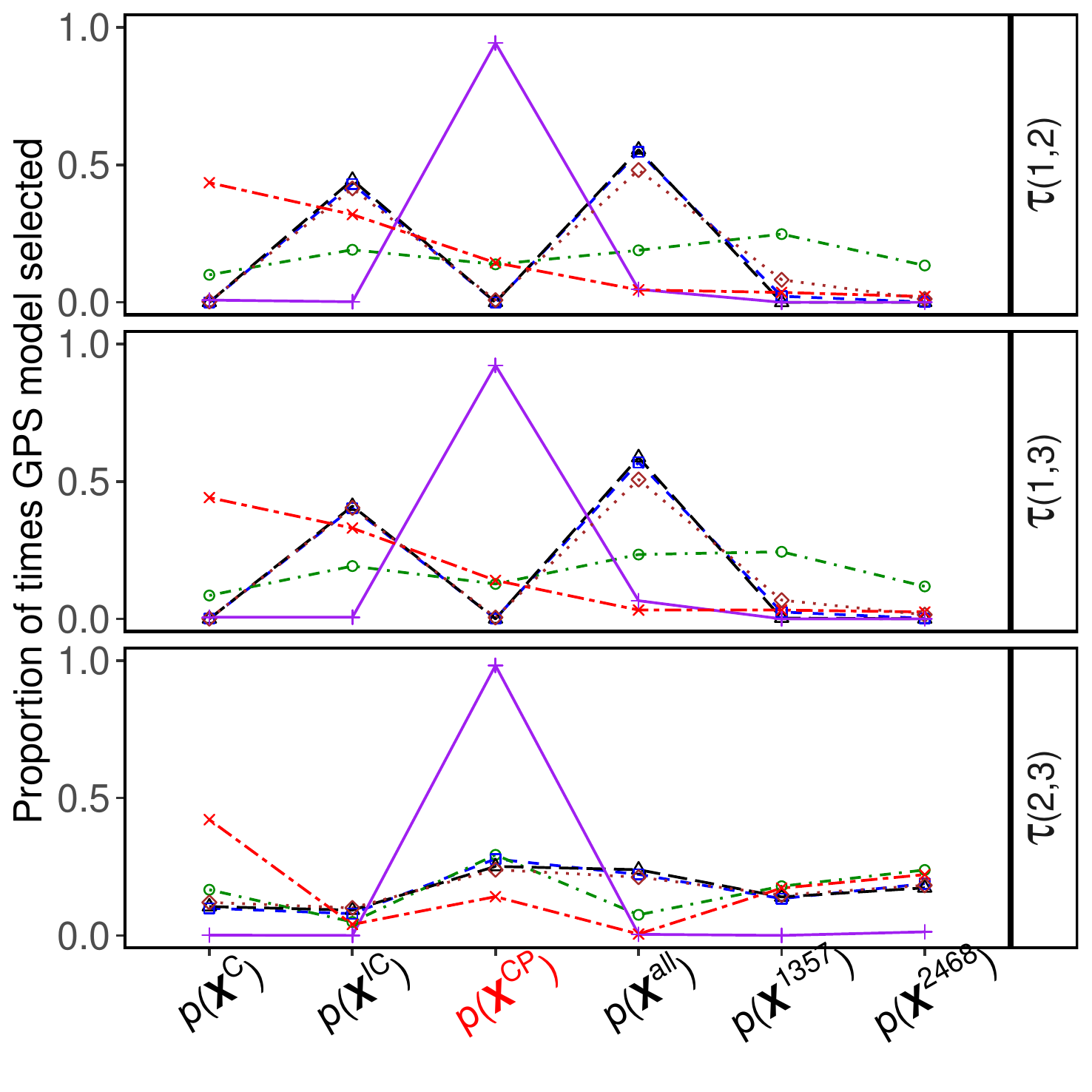}
\par\end{centering}
}
\par\end{centering}
\subfloat{\includegraphics[width=14cm,height=1.5cm]{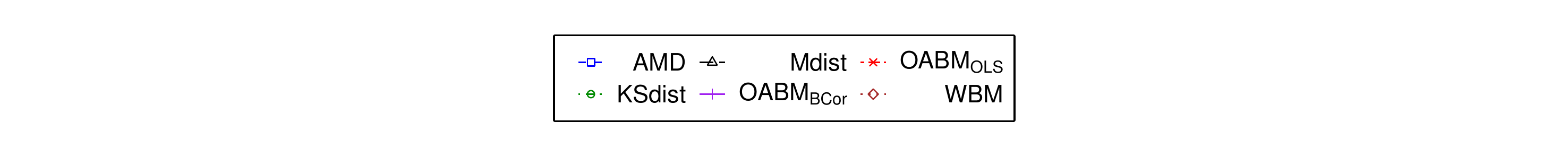}}
\caption{\label{proportionms} Proportion of benchmark models selected by the six balance measures
over 1000 simulations under 2 linear and 2 nonlinear scenarios. The optimal GPS model is colored red on the $x$-axis.} 
\end{figure}

\begin{table}[h!]
\captionsetup[subfloat]{justification=centering}
\centering
\subfloat[$u=2,v=1$\protect \\
Outcome \textit{linear} in covariates]{
\begin{tabular}{lccc}
\hline 
 & $\tau(1,2)$ & $\tau(1,3)$ & $\tau(2,3)$\tabularnewline
\hline 
$\mathtt{KSdist}$ & 0.586 & 0.688 & 0.725\tabularnewline
$\mathtt{Mdist}$ & 0.908 & 0.907 & 0.929\tabularnewline
$\mathtt{AMD}$ & 0.897 & 0.900 & 0.911\tabularnewline
$\mathtt{WBM}$ & 0.958 & 0.963 & 0.965\tabularnewline
$\mathtt{OABM}_{\mathtt{OLS}}$ & 0.964 & 0.935 & 0.950\tabularnewline
$\mathtt{OABM}_{\mathtt{BCor}}$ & 0.954 & 0.936 & 0.948\tabularnewline
\hline 
\end{tabular}
}
\subfloat[\label{s2t3 linear}$u=2,v=2$\protect \\
Outcome \textit{linear} in covariates]{
\begin{tabular}{lccc}
\hline 
 & $\tau(1,2)$ & $\tau(1,3)$ & $\tau(2,3)$\tabularnewline
\hline 
$\mathtt{KSdist}$ & 0.744 & 0.800 & 0.851\tabularnewline
$\mathtt{Mdist}$ & 0.932 & 0.898 & 0.907\tabularnewline
$\mathtt{AMD}$ & 0.918 & 0.887 & 0.896\tabularnewline
$\mathtt{WBM}$ & 0.986 & 0.945 & 0.950\tabularnewline
$\mathtt{OABM}_{\mathtt{OLS}}$ & 0.987 & 0.917 & 0.917\tabularnewline
$\mathtt{OABM}_{\mathtt{BCor}}$ & 0.969 & 0.916 & 0.916\tabularnewline
\hline 
\end{tabular}
}

\subfloat[$u=2,v=1$\protect \\
Outcome \textit{nonlinear} in covariates]{
    \begin{tabular}{lccc}
    \hline 
     & $\tau(1,2)$ & $\tau(1,3)$ & $\tau(2,3)$\tabularnewline
    \hline 
    $\mathtt{KSdist}$ & 0.929 & 0.939 & 0.937\tabularnewline
    $\mathtt{Mdist}$ & 0.914 & 0.904 & 0.898\tabularnewline
    $\mathtt{AMD}$ & 0.908 & 0.905 & 0.900\tabularnewline
    $\mathtt{WBM}$ & 0.900 & 0.909 & 0.890\tabularnewline
    $\mathtt{OABM}_{\mathtt{OLS}}$ & 0.930 & 0.937 & 0.946\tabularnewline
    $\mathtt{OABM}_{\mathtt{BCor}}$ & 0.929 & 0.938 & 0.934\tabularnewline
    \hline 
    \end{tabular}
}
\subfloat[$u=2,v=2$\protect \\
Outcome \textit{nonlinear} in covariates]{
\begin{tabular}{lccc}
\hline 
 & $\tau(1,2)$ & $\tau(1,3)$ & $\tau(2,3)$\tabularnewline
\hline 
$\mathtt{KSdist}$ & 0.941 & 0.942 & 0.932\tabularnewline
$\mathtt{Mdist}$ & 0.905 & 0.908 & 0.900\tabularnewline
$\mathtt{AMD}$ & 0.903 & 0.911 & 0.903\tabularnewline
$\mathtt{WBM}$ & 0.898 & 0.920 & 0.900\tabularnewline
$\mathtt{OABM}_{\mathtt{OLS}}$ & 0.927 & 0.918 & 0.935\tabularnewline
$\mathtt{OABM}_{\mathtt{BCor}}$ & 0.941 & 0.942 & 0.946\tabularnewline
\hline 
\end{tabular}
}
\caption{\label{tab:Coverage-of-asymptotic}Coverage rates of asymptotic 95\%
confidence intervals under 2 linear and 2 nonlinear scenarios.}
\end{table}

Figure \ref{fig:boxplots} summarizes the performance of the GPSM
estimator based on the six benchmark models and five measures under
the 4 chosen scenarios. Models $\mathtt{p(\bm{X}^{1357})}$ and $\mathtt{p(\bm{X}^{2468})}$
both yield biased estimates in all four scenarios due to failure to
include all confounders. While all other benchmark models lead to
unbiased estimates, estimates by matching on $\mathtt{p(\bm{X}^{CP})}$ result
in the smallest variance. Under the linear outcome design, $\mathtt{OABM}_{\mathtt{OLS}}$
has mean squared error closest to the optimal benchmark model, and
outperforms all other measures. $\mathtt{OABM}_{\mathtt{BCor}}$ results in mean squared error smaller than $\mathtt{Mdist}$ and $\mathtt{AMD}$, and performs
similar to $\mathtt{WBM}$. Bias occurs when matching on GPS models
selected based on $\mathtt{KSdist}$. When the mean potential outcome is nonlinear in
the covariates, $\mathtt{OABM}_{\mathtt{OLS}}$ no longer dominates in performance
due to the outcome model misspecification. In this case, because of
the model-free property of the ball correlation, $\mathtt{OABM}_{\mathtt{BCor}}$
shows a slight advantage over all other measures in terms of MSE.
For interval estimation, Table \ref{tab:Coverage-of-asymptotic} shows
that coverage rates for nominal $95\%$ confidence intervals for $\mathtt{OABM}_{\mathtt{BCor}}$
do not deviate much from the specified $95\%$ probability in all
four scenarios; Similarly, $\mathtt{OABM}_{\mathtt{OLS}}$ has coverage rates
close to the specified $95\%$ when each mean potential outcome is a linear function of
the covariates.

In Figure \ref{proportionms} we present the proportion
of times each of the six benchmark GPS models are selected by the
measures. When outcome is linear in covariates, $\mathtt{OABM}_{\mathtt{OLS}}$
consistently selects $\mathtt{p(\bm{X}^{CP})}$, the GPS model that results
in the smallest mean squared error. Following $\mathtt{OABM}_{\mathtt{OLS}}$,
$\mathtt{OABM}_{\mathtt{BCor}}$ selects $\mathtt{p(\bm{X}^{CP})}$ with high proportions,
while occasionally selecting the benchmark model that includes all covariates.
$\mathtt{Mdist}$ and $\mathtt{AMD}$ tend to select GPS models that
include the IVs, thereby explaining their large estimation variance.
When the potential outcome is nonlinear in the covariates, $\mathtt{OABM}_{\mathtt{OLS}}$
no longer selects $\mathtt{p(\bm{X}^{CP})}$, while $\mathtt{OABM}_{\mathtt{BCor}}$ is
still able to identify it thanks to the model-free property of the
ball correlation. In this case, in addition to $\mathtt{Mdist}$ and
$\mathtt{AMD}$, $\mathtt{WBM}$ also frequently selects GPS models
that include the instrumental variables. $\mathtt{KSdist}$'s worst
performance in estimation could be explained by its inability to rule
out the two biased benchmark models.

\section{An Application}
\label{sec6}

We apply the model selection method to the Tutoring dataset from the \textit{TriMatch} R package (\citealp{jason2017}), which
contains results from a study examining the effects of tutoring services
on course grades. A total of 1172 observations consist of 918 students
who did not receive any form of tutoring (Control), 134 students who
received the first form of tutoring (Treat1), and 90 students who
received the second form of tutoring (Treat2). The course grade the
student earn is the outcome variable and takes on one of five numeric
values: 4=A, 3=B, 2=C, 1=D, 0=F or W. Information on gender, ethnicity,
military service status of the student, non-native English speaker
status, education level of the student's mother, education
level of the student's father, age of the student,
employment status, household income, number of transfer credits, and
grade point average are collected as baseline covariates. The
objective in this analysis is to posit and select the best generalized
propensity score models using the outcome-adjusted balance measure. The fitted
generalized propensity scores (probabilities of a student receiving each of the
three tutoring services) will serve as matching variables for the GPSM
estimator to estimate and make inference about the pairwise average treatment effects. 

We add all first order interaction and second order terms to the original
11 covariates to form a total of 84 covariates. We apply the group
lasso (\citealp{yuan2006}) and select 10 covariates
that are strong predictors for tutoring service assignment. We also
apply the lasso and selected 26 covariates for predicting $Y(\text{Control})$,
11 for predicting $Y(\text{Treat1})$, and 4 for predicting $Y(\text{Treat2})$.
For estimation of $\mathbb{E}\left\{ Y(\text{Control})\right\} $,
we tentatively categorize the relevant covariates into 6 instrumental
variables, 4 confounders, and 22 precision variables based on the
lasso variable selection results. We follow the same procedure to
partition the relevant covariates for estimation of $\mathbb{E}\left\{ Y(\text{Treat1})\right\} $
into 10 instrumental variables, 0 confounders, and lastly 11 precision
variables, and 10 instrumental variables, 0 confounders, and 4 precision
variables for estimating $\mathbb{E}\left\{ Y(\text{Treat2})\right\} $.

For each treatment level $w$, we estimate the average potential outcome
$\mathbb{E}\left\{ Y(w)\right\} $ by positing the following three
candidate GPS models:
\begin{itemize}
\item Model 1: multinomial logit with all relevant covariates entering linearly
\item Model 2: multinomial logit with confounders and precision variables
entering linearly
\item Model 3: multinomial logit with confounders and top 3 precision variables
ranked by lasso coefficients entering linearly
\end{itemize}
We consider two versions of the outcome-adjusted balance measure to conduct GPS model selection.
First, $\mathtt{OABM}_{\mathtt{BCor}}$ uses the ball correlation to construct
the penalty weights in the outcome-adjusted balance measure; it represents our lack
of confidence in describing the true relationship between the full
set of covariates and the course grades conditional on the choice
of tutoring service. On the other hand, since the outcome is ordinal,
the ordinal probit model ($\mathtt{OABM}_{\mathtt{OP}}$) could serve as a reasonable
approximation to the underlying relationship between the covariates
and the outcome conditional on the treatment assignment. We evaluate
the two balance measures on the three proposed models for each separate
average potential outcome estimation, and the model selection results
are shown in Table \ref{tab:realdat-model-selection}. For estimating
$\mathbb{E}\left\{ Y(\text{Treat1})\right\} $ and $\mathbb{E}\left\{ Y(\text{Treat2})\right\} $,
two variants of $\mathtt{OABM}_{\rho}$ both select Model 3. For estimating
$\mathbb{E}\left\{ Y(\text{Control})\right\} $, $\mathtt{OABM}_{\mathtt{BCor}}$
picks the most conservative Model 1, while $\mathtt{OABM}_{\mathtt{OP}}$ selects
Model 3. The selected model specifications are subsequently used
to fit the data and compute the estimated generalized propensity scores, which
serve as matching variables for the GPSM estimator to produce point estimates and 95\% confidence intervals for the pairwise ATEs, as shown in Table \ref{tab:realdat-Pairwise-differences-means}.
Since none of the confidence intervals contains zero, there is sufficient statistical evidence that students undergoing the second form of tutoring will have better course grades than if they receive the first form of tutoring or no tutoring at all. Receiving the first form of tutoring also seems to help increase course grades compared to not receiving any form of tutoring at all.

\begin{table}[h!]
\begin{centering}
\begin{tabular}{ccccccc}
\hline 
 & \multicolumn{2}{c}{$\mathbb{E}[Y(Control)]$} & \multicolumn{2}{c}{$\mathbb{E}[Y(Treat1)]$} & \multicolumn{2}{c}{$\mathbb{E}[Y(Treat2)]$}\tabularnewline
\cline{2-7} \cline{3-7} \cline{4-7} \cline{5-7} \cline{6-7} \cline{7-7} 
Model & $\mathtt{OABM}_{\mathtt{BCor}}$ & $\mathtt{OABM}_{\mathtt{OP}}$ & $\mathtt{OABM}_{\mathtt{BCor}}$ & $\mathtt{OABM}_{\mathtt{OP}}$ & $\mathtt{OABM}_{\mathtt{BCor}}$ & $\mathtt{OABM}_{\mathtt{OP}}$\tabularnewline
\hline 
1 & \textcolor{red}{115.69} & 5.12 & 173.38 & 42.36 & 108.99 & 11.07\tabularnewline
2 & 197.76 & 9.11 & 71.83 & 3.46 & 18.64 & 43.50\tabularnewline
3 & \textcolor{black}{222.08} & \textcolor{orange}{1.58} & \textcolor{red}{26.01} & \textcolor{orange}{2.79} & \textcolor{red}{33.65} & \textcolor{orange}{5.20}\tabularnewline
\hline 
\end{tabular}
\par\end{centering}
\caption{\label{tab:realdat-model-selection}GPS model selection results; the
smallest balance measures based on $\mathtt{OABM}_{\mathtt{BCor}}$ are highlighted
in red, and those based on $\mathtt{OABM}_{\mathtt{OP}}$ are highlighted in
orange.}
\end{table}
\begin{table}[h!]
\begin{centering}
\begin{tabular}{cccc}
\hline 
Measure & $\widehat{\mathbb{E}}\left\{ Y(\text{Treat1})-Y(\text{Control})\right\} $ & $\widehat{\mathbb{E}}\left\{ Y(\text{Treat2})-Y(\text{Control})\right\} $ & $\widehat{\mathbb{E}}\left\{ Y(\text{Treat2})-Y(\text{Treat1})\right\} $\tabularnewline
\hline 
$\mathtt{OABM}_{\mathtt{BCor}}$ & 0.35(0.14, 0.56) & 0.66(0.47, 0.85) & 0.31(0.08, 0.55)\tabularnewline
$\mathtt{OABM}_{\mathtt{OP}}$ & 0.31(0.10, 0.52) & 0.63(0.46, 0.81) & 0.33(0.09, 0.56)\tabularnewline
\hline 
\end{tabular}
\par\end{centering}
\caption{\label{tab:realdat-Pairwise-differences-means}Pairwise ATE estimates
(95\% CI)}
\end{table}

\section{Discussion}\label{sec7}

In this article, we present a novel balance measure that can
select the optimal generalized propensity score model from a set
of candidate models for the estimation of each average potential outcome. Given a set of candidate GPS models, the outcome-adjusted balance
measure works as follows. First we impute all measured covariates
by matching on each candidate GPS model as if they were missing. For each
candidate model, the balance measure then evaluates a weighted sum of the absolute mean
differences between the imputed covariates and their original sample
values. These absolute differences are each scaled by a penalty,
which takes into account the covariate's relationship with the potential
outcome to ensure that only the desired covariates are selected into
the GPS model. Under certain conditions, we theoretically show that in large samples, the GPS specification that minimizes the proposed balance measure is the optimal GPS model. As a result, the GPSM estimator based on a GPS model selected by the outcome-adjusted balance measure is consistent and efficient for the pairwise ATEs.

Since the proposed balance measure incorporates outcome information to conduct GPS model selection, this is contradictory to Rubin's principle that modeling of the assignment mechanism should be carried out at the design stage, without accessing any outcomes \citep{rubin2007design}. However, for efficiency considerations, leveraging outcome information is necessary. In practice, if pilot studies exist, the outcome information can be borrowed from existing studies.

In addition to limitations shared by all (G)PS based methods such
as the existence of unmeasured confounders, we review some possible
limitations specific to the outcome-adjusted balance measure. First, as discussed in previous sections, when the outcome model cannot be well characterized, model selection consistency for the outcome-adjusted balance measure based on the ball correlation only holds approximately, depending on how large the collider bias is. Similarly, $\mathtt{OABM_{BCor}}$ is also unable to exclude a noise variable that is a collider of instrumental variables and precision variables, a scenario commonly known as the M-Bias structure. Including this noise variable in the GPS model will introduce bias in estimating the ATE. Despite these theoretical anomalies, it is rarely the case that the magnitude of the collider bias described above will be substantial in practice (\citealp{ding2014}). Lastly,
the success of balance measure relies on the fact that the set of
posited GPS models must contain the optimally specified model or
at least one close approximation to it. If the set of posited GPS
models only consists of severely misspecified models, then the balance
measure may not be able to suggest the "best" one among them. 

Future research work may involve extending the current balance measure to
settings with longitudinal or time-to-event outcomes (\citealp{tang2020causal}), as well as to predictive mean matching (\citealp{yang2020}) in missing data, where the predictive mean function requires estimation and model selection. 


\section*{Acknowledgments}
Yang is partly supported by the National Science Foundation grant
DMS 1811245, National Institute of Aging grant 1R01AG066883, and National
Institute of Environmental Health Sciences 1R01ES031651.
 {\it Conflict of Interest}: None declared.

\bibliographystyle{apalike}
\bibliography{refs}

\end{document}




\begin{center}
\Large
\textbf{Supplementary Material}
\end{center}

This document contains supplementary information for the main article.
In Section S.1, we make explicit some notation and assumptions needed for
proving Lemma \ref{Lemma:Let-all-confounders} and Theorem \ref{theorem1-model selection consistency}.
Section S.2 presents the proofs. Section S.3 presents the definition of $\bm{c}(w)$ and a variance estimator
for the GPSM estimator after model selection. Section S.4 summarizes the steps we take to perform model selection using the proposed balance measure. Section S.5 presents definitions for the existing balance measures. Section S.6 contains results of the extended simulation study. 

\section{Notation and Assumptions}

We first introduce some useful notation that will appear later on
in the proof. Define the conditional mean and variance of a single
covariate $X$ given $p\left(w\mid\boldsymbol{X}\right)$:

\[
\bar{\mu}_{X}\left\{ p\left(w\mid\boldsymbol{X}\right)\right\} =\mathbb{E}\left\{ X\mid p(w\mid\boldsymbol{X})\right\} ,
\]
\[
\bar{\sigma}_{X}^{2}\left\{ p\left(w\mid\boldsymbol{X}\right)\right\} =\mathbb{V}\left\{ X\mid p(w\mid\boldsymbol{X})\right\} .
\]
Define the conditional means and variances of $Y(w)$ given covariates
$\boldsymbol{X}$ and given $p\left(w\mid\boldsymbol{X}\right)$:
\[
\mu(w,\bm{x})=\mathbb{E}\left\{ Y(w)\mid W=w,\boldsymbol{X}=\bm{x}\right\} ,
\]
\[
\bar{\mu}(w,p)=\mathbb{E}\left\{ Y(w)\mid W=w,p\left(w\mid\boldsymbol{X}\right)=p\right\} ,
\]
\[
\sigma^{2}(w,\bm{x})=\mathbb{V}\left\{ Y(w)\mid W=w,\boldsymbol{X}=\bm{x}\right\} ,
\]
\[
\bar{\sigma}^{2}(w,p)=\mathbb{V}\left\{ Y(w)\mid W=w,p\left(w\mid\boldsymbol{X}\right)=p\right\} .
\]
We let $\rho_{w,N}(X,Y)=\rho_{N}\left(X,Y\mid W=w\right)$ denote
the measure of correlation between $X$ and $Y$ conditional on $W=w$.
Let $p^{*}=p(w\mid\boldsymbol{X}^{\mathcal{C}},\boldsymbol{X}^{\mathcal{P}})$
denote the optimal generalized propensity score model. To show dependence of $\widehat{X}_{\rm gpsm}(w)$ on a GPS model $p(w|\bm{X})$ explicitly and to simplify notation, we let $\widehat{X}(w,p(w|\bm{X}))=\widehat{X}_{\rm gpsm}(w)$. For convenience, also let $\rho_{w,N}(X,Y)$ denote $\rho_{N}(X,Y|W=w)$.

\begin{Assumption} \label{martingale regularity}$p(w\mid\boldsymbol{X})$
has a continuous distribution with compact support $[\underline{p},\overline{p}]$
with a continuous density function. $\bar{\mu}_{X}(w,p)$ is Lipschitz-continuous
in $p$. For some $\delta>0$, $\mathbb{E}\{|X|^{2+\delta}|W=w,p(w\mid\boldsymbol{X})=p\} $
is uniformly bounded.

\end{Assumption}

\begin{Assumption} We have a random sample of size $N$ from a large
population.

\end{Assumption} 


We include a lemma and a theorem, which will be useful for proving
Lemma \ref{Lemma:Let-all-confounders}. The lemma is proven by \citet{abadie2016}.
Theorem \ref{martigale CLT} is the Central Limit Theorem for martingale
arrays (e.g. \citealp{billingsley1995}).

\begin{lemma}\label{lemma AI16} Suppose that $\left(W_{1},X_{1}\right),...,\left(W_{N},X_{N}\right)$
are independent and identically distributed, where the density of
$X$ is continuous on $\left[a,b\right]$, and $Pr(W=w)>0$ for $w\in\mathbb{W}$.
Assume also that $\sigma^{2}\left(w,X_{i}\right)$ is uniformly bounded.
For a given $w$, let $p^{*}=Pr(W=w)$, we have 
\[
\frac{1}{N_{w}}\sum_{i=1}^{N}D_{i}(w)\sigma^{2}\left(w,X_{i}\right)K(i,w)\stackrel{p}{\rightarrow}\mathbb{E}\left\{ \sigma^{2}\left(w,X_{i}\right)\left(\frac{p^{*}}{1-p^{*}}\right)^{1-2D_{i}(w)}\mid W_{i}=w\right\} 
\]

and
\begin{align*}
\frac{1}{N_{w}}\sum_{i=1}^{N}D_{i}(w)\sigma^{2}\left(w,X_{i}\right)K(i,w)^{2}\stackrel{p}{\rightarrow} & \mathbb{E}\left\{ \sigma^{2}\left(w,X_{i}\right)\left(\frac{p^{*}}{1-p^{*}}\right)^{1-2D_{i}(w)}\mid W_{i}=w\right\} \\
 & +\frac{3}{2}\mathbb{E}\left\{ \sigma^{2}\left(w,X_{i}\right)\left(\frac{p^{*}}{1-p^{*}}\right)^{2(1-2D_{i}(w))}\mid W_{i}=w\right\} 
\end{align*}
\end{lemma}

\begin{theorem} \label{martigale CLT}Let $X_{n,k}$, $1\leq k\leq m_{n}$
be a martingale array with respect to $F_{n,k}$ and let $S_{n,k}=\sum_{i=1}^{k}X_{n,i}$.
If $\mathbb{E}_{\max_{1\leq j\leq m_{n}}}|X_{n,j}|\rightarrow0$ or
$\sum_{j=1}^{m_{n}}\mathbb{E}|X_{n,j}|^{2+\delta}\rightarrow0$ for
some $\delta>0$ and $\sum_{j=1}^{m_{n}}X_{n,j}^{2}\stackrel{p}{\rightarrow}\sigma^{2}$,
then $S_{n,m_{n}}\stackrel{d}{\rightarrow}N(0,\sigma^{2})$.

\end{theorem}

\section{Proofs}

We first prove Lemma \ref{Lemma:Let-all-confounders}. Recall that
$D_{i}(w)$ is the indicator of whether unit $i$ received treatment
$w$, and $K(i,w)$ is the number of times unit $i$ is used as a
match when each unit is matched with replacement to one closest unit
at treatment level $w$. By definition of the matching function, we
can express the sum of imputed quantities in terms of $D_{i}(w)$,
$K(i,w)$ and the original values, and vice versa:
\begin{equation}
\sum_{i=1}^{N}X_{m\{w,p(w\mid\boldsymbol{X}_{i})\}}=\sum_{i=1}^{N}D_{i}(w)\left\{ 1+K(i,w)\right\} X_{i}\label{imputeX}
\end{equation}
\begin{equation}
\sum_{i=1}^{N}\left\{ 1-D_{i}(w)\right\} \bar{\mu}_{X}\left\{ p\left(w\mid\boldsymbol{X}_{m\{w,p(w\mid\boldsymbol{X}_{i})\}}\right)\right\} =\sum_{i=1}^{N}D_{i}(w)K(i,w)\bar{\mu}_{X}\left\{ p\left(w\mid\boldsymbol{X}_{i}\right)\right\} \label{imputeMuX}
\end{equation}
\begin{proof}[Proof of Lemma \ref{Lemma:Let-all-confounders}]
We follow a similar argument given by \citet{abadie2016}
and \citet{yang2016}. Using the two properties
shown in Equation (\ref{imputeX}) and (\ref{imputeMuX}), we obtain
\begin{align*}
 & \sqrt{N}\left(\frac{1}{N}\sum_{i=1}^{N}X_{m\{w,p(w\mid\boldsymbol{X}_{i})\}}-\overline{X}\right)\\
= & \frac{1}{\sqrt{N}}\sum_{i=1}^{N}D_{i}(w)\left\{ 1+K(i,w)\right\} \left[X_{i}-\bar{\mu}_{X}\left\{ p\left(w\mid\boldsymbol{X}_{i}\right)\right\} \right]\\
 & +\frac{1}{\sqrt{N}}\sum_{i=1}^{N}\left[\bar{\mu}_{X}\left\{ p\left(w\mid\boldsymbol{X}_{i}\right)\right\} -X_{i}\right]\\
 & +\ensuremath{\frac{1}{\sqrt{N}}\sum_{i=1}^{N}\left\{ 1-D_{i}(w)\right\} \left[\bar{\mu}_{X}\left\{ p\left(w\mid\boldsymbol{X}_{m\{w,p(w\mid\boldsymbol{X}_{i})\}}\right)\right\} -\bar{\mu}_{X}\left\{ p\left(w\mid\boldsymbol{X}_{i}\right)\right\} \right]}\\
= & \ensuremath{\frac{1}{\sqrt{N}}\sum_{i=1}^{N}\left[D_{i}(w)\left\{ 1+K(i,w)\right\} -1\right]\left[X_{i}-\bar{\mu}_{X}\left\{ p\left(w\mid\boldsymbol{X}_{i}\right)\right\} \right]}\\
 & +\ensuremath{\frac{1}{\sqrt{N}}\sum_{i=1}^{N}\left\{ 1-D_{i}(w)\right\} \left[\bar{\mu}_{X}\left\{ p\left(w\mid\boldsymbol{X}_{m\{w,p(w\mid\boldsymbol{X}_{i})\}}\right)\right\} -\bar{\mu}_{X}\left\{ p\left(w\mid\boldsymbol{X}_{i}\right)\right\} \right]}\\
= & A_{N}+R_{N}
\end{align*}
where
\[
A_{N}=\frac{1}{\sqrt{N}}\sum_{i=1}^{N}\left[D_{i}(w)\left\{ 1+K(i,w)\right\} -1\right]\left[X_{i}-\bar{\mu}_{X}\left\{ p\left(w\mid\boldsymbol{X}_{i}\right)\right\} \right]
\]
and
\[
R_{N}=\frac{1}{\sqrt{N}}\sum_{i=1}^{N}\left\{ 1-D_{i}(w)\right\} \left[\bar{\mu}_{X}\left\{ p\left(w\mid\boldsymbol{X}_{m\{w,p(w\mid\boldsymbol{X}_{i})\}}\right)\right\} -\bar{\mu}_{X}\left\{ p\left(w\mid\boldsymbol{X}_{i}\right)\right\} \right].
\]
We rewrite $A_{N}$ as
\[
A_{N}=\ensuremath{\sum_{k=1}^{N}\xi_{N,k}},
\]
where
\[
\ensuremath{\xi_{N,k}=\frac{1}{\sqrt{N}}\left[D_{k}(w)\left\{ 1+K(k,w)\right\} -1\right]\left[X_{k}-\bar{\mu}_{X}\left\{ p\left(w\mid\boldsymbol{X}_{k}\right)\right\} \right]},
\]
for $1\leq k\leq N$. Consider the $\sigma$-fields $F_{N,k}=\ensuremath{\text{\ensuremath{\sigma}}\left\{ D_{1}(w),\ldots,D_{N}(w),p\left(w\mid\boldsymbol{X}_{1}\right),\ldots,p\left(w\mid\boldsymbol{X}_{N}\right),X_{1},...,X_{k}\right\} }$
for $1\leq k\leq N$. Since we assumed that $D_{i}(w)\ \indep\ X_{i}\ \mid\ p(w\mid\boldsymbol{X}_{i})$,
we have
\begin{align*}
\mathbb{E}\left(\sum_{k=1}^{m}\xi_{N,k}\mid F_{N,m}\right)= & \sum_{k=1}^{m-1}\xi_{N,k}\\
 & +\frac{1}{\sqrt{N}}\left[D_{m}(w)\left\{ 1+K(m,w)\right\} -1\right]\left[\mathbb{E}\left\{ X_{m}\mid D_{m}(w),p\left(w\mid\boldsymbol{X}_{m}\right)\right\} -\bar{\mu}_{X}\left\{ p\left(w\mid\boldsymbol{X}_{m}\right)\right\} \right]\\
= & \sum_{k=1}^{m-1}\xi_{N,k},
\end{align*}
and so for each $N\geq1$,
\[
\ensuremath{\left\{ \sum_{j=1}^{i}\xi_{N,j},F_{N,i},1\leq i\leq N\right\} }
\]
is a martingale. To obtain the asymptotic normality of $A_{N}$, we
apply Theorem \ref{martigale CLT}. The following two conditions are
sufficient: 
\begin{equation}
\ensuremath{\sum_{k=1}^{N}\mathbb{E}\left(|\xi|_{N,k}^{2+\delta}\right)\rightarrow0}\text{ for some }\ensuremath{\delta>0}\label{lyap}
\end{equation}
and
\begin{equation}
\ensuremath{\sum_{k=1}^{N}\mathbb{E}\left(\xi_{N,k}^{2}\mid F_{N,k-1}\right)}\stackrel{p}{\rightarrow}\mathbb{E}\left[\bar{\sigma}_{X}^{2}\left\{ p\left(w\mid\boldsymbol{X}\right)\right\} \left\{ \frac{3}{2}\frac{1}{p\left(w\mid\boldsymbol{X}\right)}-\frac{1}{2}p\left(w\mid\boldsymbol{X}\right)-1\right\} \right].\label{var}
\end{equation}

Equation (\ref{lyap}) is the Lyapounov condition (which is sufficient
for the usual Lindeberg condition to hold). Because of Assumption
\ref{martingale regularity}, $\bar{\mu}_{X}\left\{ p\left(w\mid\boldsymbol{X}_{k}\right)\right\} $
are continuous on a compact support, these functions are also bounded.
Let $C_{\bar{\sigma}_{X}^{2+\delta}}$ be a bound on $E[\left|X_{i}-\bar{\mu}_{X}\left\{ W_{i},p\left(w\mid\boldsymbol{X}_{i}\right)\right\} \right|^{2+\delta}\mid W_{i},p\left(w\mid\boldsymbol{X}_{i}\right)]$
for $w\in\mathbb{W}$ and $p\in[\underline{p},\overline{p}]$. Using
the Law of Iterated Expectation and the fact that $K(k,w)$ has bounded
moments (see Lemma A.8 in \citet{abadie2016}),
we obtain
\begin{align*}
\sum_{k=1}^{N}\mathbb{E}\left[\left|\xi_{N,k}\right|^{2+\delta}\right] & =N^{-\delta/2}\mathbb{E}\left[\left|D_{k}(w)\left\{ 1+K(k,w)\right\} -1\right|^{2+\delta}\left|X_{k}-\bar{\mu}_{X}\left\{ p\left(w\mid\boldsymbol{X}_{k}\right)\right\} \right|^{2+\delta}\right]\\
 & \leq\frac{C_{\bar{\sigma}_{X}^{2+\delta}}E\left[\left\{ 1+K(k,w)\right\} ^{2+\delta}\right]}{N^{\delta/2}}\rightarrow0.
\end{align*}

To prove equation (\ref{var}) notice that 
\[
\sum_{k=1}^{N}\mathbb{E}\left(\xi_{N,k}^{2}\mid F_{N,k-1}\right)=\frac{1}{N}\sum_{k=1}^{N}\left[D_{k}(w)\left\{ 1+K(k,w)\right\} -1\right]^{2}\bar{\sigma}_{X}^{2}\left\{ p\left(w\mid\boldsymbol{X}_{k}\right)\right\} .
\]
Then, Lemma \ref{lemma AI16} implies equation (\ref{var}) after
some algebra. 

The term $R_{N}/\sqrt{N}$ is the conditional bias for covariate due
to matching discrepancy. Theorem 1 from \citet{abadie2006}
states that under appropriate conditions, the conditional bias for
potential outcome $Y(w)$ due to matching discrepancy $R_{N}(w)/\sqrt{N}=O_{p}\left(N^{-1/k}\right)$
where $k$ is the number of continuous matching variables. In our
case, $k=1$ and so we have $R_{N}=O_{p}\left(N^{-1/2}\right)$. This
is sufficient for showing $R_{N}=o_{p}(1)$. 
\end{proof}

Now we are ready to prove Theorem \ref{theorem1-model selection consistency}.

\begin{proof}[Proof of Theorem \ref{theorem1-model selection consistency}]
We first prove Part (i).
Without loss of generality, let $\bm{X}=(X^{(1)},X^{(2)},X^{(3)},X^{(4)})^{\text{T}}$
be the vector of baseline covariates, where $X^{(1)}$ is a confounder,
$X^{(2)}$ is an instrumental variable, $X^{(3)}$ is a precision
variable, and $X^{(4)}$ is a null variable. By construction, we have
$p^{*}=p(w\mid X^{(1)},X^{(3)})$. For any specification $p_{k}\in\mathcal{\mathscr{P}}$,
we want to show 
\[
\mathbb{P}\{\mathtt{OABM_{CAN}}(w,p^{*})\leq\mathtt{OABM_{CAN}}(w,p_{k})\}\rightarrow1.
\]
It suffices to show 
\begin{equation}
\mathbb{P}\left\{ \sum\limits _{j=1}^{4}\zeta_{j}(w,p_{k})\left|\widehat{X}^{(j)}(w,p_{k})-\overline{X}^{(j)}\right|\leq\sum_{i=1}^{4}\zeta_{j}(w,p^{*})\left|\widehat{X}^{(j)}(w,p^{*})-\overline{X}^{(j)}\right|\right\} \label{eq:wts}
\end{equation}
converges to $0$ as $N\rightarrow\infty$.

By Lemma \ref{Lemma:Let-all-confounders}, we can quickly obtain the
following convergence in probability results for $\mathtt{OABM_{CAN}}(w,p^{*})=A_{1}+A_{2}+A_{3}+A_{4}$,
where:
\[
A_{1}=\zeta_{1}(w,p^{*})\left|\widehat{X}^{(1)}(w,p^{*})-\overline{X}^{(1)}\right|=\left|\widehat{X}^{(1)}(w,p^{*})-\overline{X}^{(1)}\right|/\rho_{w,N}(X^{(1)},Y)\overset{p}{\rightarrow}0
\]

\[
A_{2}=\zeta_{2}(w,p^{*})\left|\widehat{X}^{(2)}(w,p^{*})-\overline{X}^{(2)}\right|=N^{1/3}\rho_{w,N}(X^{(2)},Y)\left|\widehat{X}^{(2)}(w,p^{*})-\overline{X}^{(2)}\right|\overset{p}{\rightarrow}0
\]

\[
A_{3}=\zeta_{3}(w,p^{*})\left|\widehat{X}^{(3)}(w,p^{*})-\overline{X}^{(3)}\right|=\left|\widehat{X}^{(3)}(w,p^{*})-\overline{X}^{(3)}\right|/\rho_{w,N}(X^{(3)},Y)\overset{p}{\rightarrow}0
\]
\[
A_{4}=\zeta_{4}(w,p^{*})\left|\widehat{X}^{(4)}(w,p^{*})-\overline{X}^{(4)}\right|=N^{1/3}\rho_{w,N}(X^{(4)},Y)\left|\widehat{X}^{(4)}(w,p^{*})-\overline{X}^{(4)}\right|\overset{p}{\rightarrow}0.
\]
The rest of the proof is divided into four parts, with each part corresponding
to a type of misspecified model $p_{k}\neq p^{*}$.

\textbf{Part 1: (Penalize the inclusion of IV)}
Consider model $p_{k}=p(w\mid X^{(1)},X^{(2)},X^{(3)})$, which results
from adding the instrumental variable to $p^{*}$. Then $\mathtt{OABM_{CAN}}(w,p_{k})=B_{1}+B_{2}+B_{3}+B_{4}$
where
\[
B_{1}=\zeta_{1}(w,p_{k})\left|\widehat{X}^{(1)}(w,p_{k})-\overline{X}^{(1)}\right|=\left|\widehat{X}^{(1)}(w,p_{k})-\overline{X}^{(1)}\right|/\rho_{w,N}(X^{(1)},Y)\overset{p}{\rightarrow}0
\]
\[
B_{3}=\zeta_{3}(w,p_{k})\left|\widehat{X}^{(3)}(w,p_{k})-\overline{X}^{(3)}\right|=\left|\widehat{X}^{(3)}(w,p_{k})-\overline{X}^{(3)}\right|/\rho_{w,N}(X^{(3)},Y)\overset{p}{\rightarrow}0
\]
\[
B_{4}=\zeta_{4}(w,p_{k})\left|\widehat{X}^{(4)}(w,p_{k})-\overline{X}^{(4)}\right|=N^{1/3}\rho_{w,N}(X^{(4)},Y)\left|\widehat{X}^{(4)}(w,p_{k})-\overline{X}^{(4)}\right|\overset{p}{\rightarrow}0,
\]
with the exception of
\[
B_{2}=\zeta_{2}(w,p_{k})\left|\widehat{X}^{(2)}(w,p_{k})-\overline{X}^{(2)}\right|=\left|\widehat{X}^{(2)}(w,p_{k})-\overline{X}^{(2)}\right|/\rho_{w,N}(X^{(2)},Y).
\]
We rewrite (\ref{eq:wts}) as 
\begin{align}
 & \mathbb{P}\left\{ B_{2}\leq o_{p}(1)\right\} \nonumber \\
= & \mathbb{P}\left\{ \left|\widehat{X}^{(2)}(w,p_{k})-\overline{X}^{(2)}\right|/\rho_{w,N}(X^{(2)},Y)\leq o_{p}(1)\right\} \nonumber \\
= & \mathbb{P}\left\{ \sqrt{N}\left|\widehat{X}^{(2)}(w,p_{k})-\overline{X}^{(2)}\right|-\sqrt{N}\rho_{w,N}(X^{(2)},Y)o_{p}(1)\leq0\right\} .\label{eq:iv}
\end{align}
By Lemma \ref{Lemma:Let-all-confounders} and the continuous mapping
theorem, we know that the first term
\[
\sqrt{N}\left|\widehat{X}^{(2)}(w,p_{k})-\overline{X}^{(2)}\right|\overset{d}{\rightarrow}\text{folded normal}.
\]
By Slutsky's theorem, we have
that the second term
\[
\sqrt{N}\rho_{w,N}(X^{(2)},Y)o_{p}(1)\overset{p}{\rightarrow}0.
\]
Therefore (\ref{eq:iv})$\rightarrow0$ as $N$ tends to infinity,
implying that the balance measure will discourage the inclusion of
IV in the GPS.

\textbf{Part 2: (Penalize the inclusion of null variable) }Proof for
the case $p_{k}=p(w\mid X^{(1)},X^{(3)},X^{(4)})$ is almost identical
to part 1.

\textbf{Part 3: (Penalize the exclusion of precision variable)}
Consider model $p_{k}=p(w\mid X^{(1)})$, which results from dropping
the precision variable from $p^{*}$. Then $\mathtt{OABM_{CAN}}(w,p_{k})=C_{1}+C_{2}+C_{3}+C_{4}$
where
\[
C_{1}=\zeta_{1}(w,p_{k})\left|\widehat{X}^{(1)}(w,p_{k})-\overline{X}^{(1)}\right|=\left|\widehat{X}^{(1)}(w,p_{k})-\overline{X}^{(1)}\right|/\rho_{w,N}(X^{(1)},Y)\overset{p}{\rightarrow}0
\]
\[
C_{2}=\zeta_{2}(w,p_{k})\left|\widehat{X}^{(2)}(w,p_{k})-\overline{X}^{(2)}\right|=N^{1/3}\rho_{w,N}(X^{(2)},Y)\left|\widehat{X}^{(2)}(w,p_{k})-\overline{X}^{(2)}\right|\overset{p}{\rightarrow}0
\]
\[
C_{3}=\zeta_{3}(w,p_{k})\left|\widehat{X}^{(3)}(w,p_{k})-\overline{X}^{(3)}\right|=N^{1/3}\rho_{w,N}(X^{(3)},Y)\left|\widehat{X}^{(3)}(w,p_{k})-\overline{X}^{(3)}\right|\overset{p}{\rightarrow}0
\]
\[
C_{4}=\zeta_{4}(w,p_{k})\left|\widehat{X}^{(4)}(w,p_{k})-\overline{X}^{(4)}\right|=N^{1/3}\rho_{w,N}(X^{(4)},Y)\left|\widehat{X}^{(4)}(w,p_{k})-\overline{X}^{(4)}\right|\overset{p}{\rightarrow}0.
\]
Now multiplying $N^{1/6}$ to both sides of the inequality in (\ref{eq:wts})
we get
\begin{align}
 & \mathbb{P}\left\{ N^{1/6}C_{3}\leq o_{p}(1)\right\} \nonumber \\
= & \mathbb{P}\left\{N^{1/2}\rho_{w,N}(X^{(3)},Y)\left|\widehat{X}^{(3)}(w,p_{k})-\overline{X}^{(3)}\right|\leq o_{p}(1)\right\} .\label{eq:pv}
\end{align}
This is because the remaining terms $A_{1},...,A_{4},C_{1},...,C_{3}$,
after multiplying by $N^{1/6}$, still converge in probability to
0. By Lemma \ref{Lemma:Let-all-confounders}, continuous mapping theorem,
consistency of $\rho_{w,N}(X^{(3)},Y)$ and Slutsky's theorem, we
get
\[
N^{1/2}\rho_{w,N}(X^{(3)},Y)\left|\widehat{X}^{(3)}(w,p_{k})-\overline{X}^{(3)}\right|\overset{d}{\rightarrow}\text{folded normal}.
\]
Therefore, (\ref{eq:pv})$\rightarrow0$ as $N$ tends to infinity,
implying that the balance measure will discourage the exclusion of
precision variables in the GPS.

\textbf{Part 4: (Penalize the exclusion of confounder)}
Finally, consider $p_{k}=p(w\mid X^{(3)})$, which results from dropping
the confounder from $p^{*}$. Then $\mathtt{OABM_{CAN}}(w,p_{k})=D_{1}+D_{2}+D_{3}+D_{4}$
where
\[
D_{2}=\zeta_{2}(w,p_{k})\left|\widehat{X}^{(2)}(w,p_{k})-\overline{X}^{(2)}\right|=N^{1/3}\rho_{w,N}(X^{(2)},Y)\left|\widehat{X}^{(2)}(w,p_{k})-\overline{X}^{(2)}\right|\overset{p}{\rightarrow}0
\]
\[
D_{3}=\zeta_{3}(w,p_{k})\left|\widehat{X}^{(3)}(w,p_{k})-\overline{X}^{(3)}\right|=\left|\widehat{X}^{(3)}(w,p_{k})-\overline{X}^{(3)}\right|/\rho_{w,N}(X^{(3)},Y)\overset{p}{\rightarrow}0
\]
\[
D_{4}=\zeta_{4}(w,p_{k})\left|\widehat{X}^{(4)}(w,p_{k})-\overline{X}^{(4)}\right|=N^{1/3}\rho_{w,N}(X^{(4)},Y)\left|\widehat{X}^{(4)}(w,p_{k})-\overline{X}^{(4)}\right|\overset{p}{\rightarrow}0,
\]
with the exception of 
\[
D_{1}=\zeta_{1}(w,p_{k})\left|\widehat{X}^{(1)}(w,p_{k})-\overline{X}^{(1)}\right|=N^{1/3}\rho_{w,N}(X^{(1)},Y)\left|\widehat{X}^{(1)}(w,p_{k})-\overline{X}^{(1)}\right|.
\]
The term $D_{1}$ is strictly positive and does not converge. Therefore
$\mathbb{P}\left\{ D_{1}\leq o_{p}(1)\right\} \rightarrow0$ as $N\rightarrow\infty$,
which implies that the balance measure will discourage the exclusion
of confounders in the GPS. 

Any other the GPS model specification (e.g. $p_{k}=p(w\mid X^{(2)},X^{(3)})$)
can be represented as a combination of the four types of misspecification.
In such cases, the result can be proven using similar techniques.\\

We now prove Part (ii).
Without loss of generality, let $\bm{X}=(X^{(1)},X^{(2)},X^{(3)},X^{(4)})$
be the baseline covariates, where $X^{(1)}$ is a confounder, $X^{(2)}$
is an instrumental variable, $X^{(3)}$ is a precision variable, and
$X^{(4)}$ is a null variable. Again, we want to show 
\[
\mathbb{P}\{\mathtt{OABM_{BCor}}(w,p_{k})\leq\mathtt{OABM_{BCor}}(w,p^{*})\}\rightarrow0.
\]

By the properties of ball correlation and Lemma \ref{Lemma:Let-all-confounders}, we can quickly obtain the following convergence in probability
results for $\mathtt{OABM_{BCor}}(w,p^{*})=A_{1}+A_{2}+A_{3}+A_{4}$,
where
\[
A_{1}=BCor_{w,N}^{-1}(X^{(1)},Y)\left|\widehat{X}^{(1)}(w,p^{*})-\overline{X}^{(1)}\right|\overset{p}{\rightarrow}0
\]
\[
A_{2}=N^{1/3}BCor_{w,N}(X^{(2)},Y)\left|\widehat{X}^{(2)}(w,p^{*})-\overline{X}^{(2)}\right|\overset{p}{\rightarrow}0
\]
\[
A_{3}=BCor_{w,N}^{-1}(X^{(3)},Y)\left|\widehat{X}^{(3)}(w,p^{*})-\overline{X}^{(3)}\right|\overset{p}{\rightarrow}0
\]
\[
A_{4}=N^{1/3}BCor_{w,N}(X^{(4)},Y)\left|\widehat{X}^{(4)}(w,p^{*})-\overline{X}^{(4)}\right|\overset{p}{\rightarrow}0
\]
up to a multiplicative constant.

\textbf{Part 1: (Exclusion of IV)}
Consider model $p_k=p(w|X^{(1)},X^{(2)},X^{(3)}) $, which
results from adding instrumental variable to the optimal model $p^{*}$.
Then $\mathtt{OABM_{BCor}}(w,p_k)=B_{1}+B_{2}+B_{3}+B_{4}$ where
\[
B_{1}=BCor_{w,N}^{-1}(X^{(1)},Y)\left|\widehat{X}^{(1)}(w,p_k)-\overline{X}^{(1)}\right|\overset{p}{\rightarrow}0
\]
\[
B_{3}=BCor_{w,N}^{-1}(X^{(3)},Y)\left|\widehat{X}^{(3)}(w,p_k)-\overline{X}^{(3)}\right|\overset{p}{\rightarrow}0
\]
\[
B_{4}=N^{1/3}BCor_{w,N}(X^{(4)},Y)\left|\widehat{X}^{(4)}(w,p_k)-\overline{X}^{(4)}\right|\overset{p}{\rightarrow}0,
\]
with the exception of
\[
B_{2}=BCor_{w,N}^{-1}(X^{(2)},Y)\left|\widehat{X}^{(2)}(w,p_k)-\overline{X}^{(2)}\right|.
\]
We can rewrite (\ref{eq:wts}) as 
\begin{align}
 & \mathbb{P}\left\{B_{2}\leq o_{p}(1)\right\}\nonumber \\
= & \mathbb{P}\left\{\frac{\sqrt{BCov_{w,N}(X^{(2)},X^{(2)})BCov_{w,N}(Y,Y)}}{BCov_{w,N}(X^{(2)},Y)}\left|\widehat{X}^{(2)}(w,p_k)-\overline{X}^{(2)}\right|\leq o_{p}(1)\right\}\nonumber \\
= & \mathbb{P}\left\{\sqrt{N}\sqrt{BCov_{w,N}(X^{(2)},X^{(2)})BCov_{w,N}(Y,Y)}\left|\widehat{X}^{(2)}(w,p_k)-\overline{X}^{(2)}\right|\leq\sqrt{N}BCov_{w,N}(X^{(2)},Y)o_{p}(1)\right\}. \label{S.8}
\end{align}
Since $BCov_{w,N}(X^{(2)},X^{(2)})BCov_{w,N}(Y,Y)\overset{p}{\rightarrow}BCov(X^{(2)},X^{(2)})BCov(Y,Y)$, by Lemma \ref{Lemma:Let-all-confounders} and continuous mapping theorem, we have
\[
\sqrt{N}\sqrt{BCov_{w,N}(X^{(2)},X^{(2)})BCov_{w,N}(Y,Y)}\left|\widehat{X}^{(2)}(w,p_k)-\overline{X}^{(2)}\right|\overset{d}{\rightarrow}\text{folded normal}.
\]
By assumption, we also have $\sqrt{N}BCov_{w,N}(X^{(2)},Y)=o_{p}(1)$.
Therefore (\ref{S.8})$\rightarrow0$ as $N$ tends to infinity, implying that
the instrumental variable will be excluded from the PS model.

\textbf{Part 2: (Exclusion of null variable) }Proof for the case $p_k=p(w|X^{(1)},X^{(2)},X^{(4)})$
is almost identical to that for part 1.

\textbf{Part 3: (Inclusion of precision variable)}
Consider model $p_k=p(w|X^{(1)})$, which results from
dropping the precision variable from the optimal model $p^{*}$.
Then $\mathtt{OABM_{BCor}}(w,p_k)=C_{1}+C_{2}+C_{3}+C_{4}$ where
\[
C_{1}=BCor_{w,N}^{-1}(X^{(1)},Y)\left|\widehat{X}^{(1)}(w,p_k)-\overline{X}^{(1)}\right|\overset{p}{\rightarrow}0
\]
\[
C_{2}=N^{1/3}BCor_{w,N}(X^{(2)},Y)\left|\widehat{X}^{(2)}(w,p_k)-\overline{X}^{(2)}\right|\overset{p}{\rightarrow}0
\]
\[
C_{3}=N^{1/3}BCor_{w,N}(X^{(3)},Y)\left|\widehat{X}^{(3)}(w,p_k)-\overline{X}^{(3)}\right|\overset{p}{\rightarrow}0
\]
\[
C_{4}=N^{1/3}BCor_{w,N}(X^{(4)},Y)\left|\widehat{X}^{(4)}(w,p_k)-\overline{X}^{(4)}\right|\overset{p}{\rightarrow}0,
\]
again up to a multiplicative constant.
Now multiplying $N^{1/6}$ to both sides of the inequality in (\ref{eq:wts}), we
get
\begin{align}
 & \mathbb{P}\left\{N^{1/6}C_{3}\leq o_{p}(1)\right\}\nonumber \\
= & \mathbb{P}\left\{\sqrt{N}BCor_{w,N}(X^{(3)},Y)\left|\widehat{X}^{(3)}(w,p_k)-\overline{X}^{(3)}\right|\leq o_{p}(1)\right\}. \label{S.9}
\end{align}
This is because the remaining terms $A_{1},...,A_{4},C_{1},C_{2},C_{4}$,
after multiplying by $N^{1/6}$, still converge in probability to
0. By Lemma \ref{Lemma:Let-all-confounders} and continuous mapping theorem again, we know
\[
\sqrt{N}BCor_{w,N}(X^{(3)},Y)\left|\widehat{X}^{(3)}(w,p_k)-\overline{X}^{(3)}\right|\overset{d}{\rightarrow}\text{folded normal}.
\]
Therefore, (\ref{S.9})$\rightarrow0$ as $N$ tends to infinity, implying
that the precision variable will be included in the PS model.

\textbf{Part 4: (Inclusion of confounder)}
Consider model $p_k=p(w|X^{(3)})$, which results from
dropping the confounder from the optimal model $p^{*}$.
Then $\mathtt{OABM_{BCor}}(w,p_k)=B_{1}+B_{2}+B_{3}+B_{4}$ where
\[
D_{2}=N^{1/3}BCor_{w,N}(X^{(2)},Y)\left|\widehat{X}^{(2)}(w,p_k)-\overline{X}^{(2)}\right|\overset{p}{\rightarrow}0
\]
\[
D_{3}=BCor_{w,N}^{-1}(X^{(3)},Y)\left|\widehat{X}^{(3)}(w,p_k)-\overline{X}^{(3)}\right|\overset{p}{\rightarrow}0
\]
\[
D_{4}=N^{1/3}BCor_{w,N}(X^{(4)},Y)\left|\widehat{X}^{(4)}(w,p_k)-\overline{X}^{(4)}\right|\overset{p}{\rightarrow}0,
\]
with the exception of 
\[
D_{1}=N^{1/3}BCor_{w,N}(X^{(1)},Y)\left|\widehat{X}^{(1)}(w,p_k)-\overline{X}^{(1)}\right|.
\]
The term $D_{1}$ is strictly positive and does not converge.
Therefore $\mathbb{P}\{D_{1}\leq o_{p}(1)\}\rightarrow0$ as $N\rightarrow\infty$,
which implies that the confounder will be included in the PS model. 

Any other the GPS model specification (e.g. $p_{k}=p(w\mid X^{(2)},X^{(3)})$)
can be represented as a combination of the four types of misspecification.
In such cases, the result can be proven using similar techniques.
\end{proof}


\section{Definition of $\bm{c}(w)$ \& Variance Estimation}
For any parametric model of the optimal GPS, suppose $\boldsymbol{X}^{\mathcal{C}\cup\mathcal{P}}$ consists of $b$ covariates. We let $p_{\bm{\beta}_{w'}}(w)$ be such that 
\[
\frac{\partial p}{\partial \bm{\beta}_{w'}}(w\mid\bm{X}^{\mathcal{C}\cup\mathcal{P}}=\bm{x};\bm{\beta})=\bm{x}\times p_{\bm{\beta}_{w'}}(w|\bm{x};\bm{\beta}),
\]and define
\[
\begin{aligned}\bm{c}\left(w\right)= & \mathbb{E}\left[\operatorname{cov}\left\{ \boldsymbol{X}^{\mathcal{C}\cup\mathcal{P}},\mu\left(w,\boldsymbol{X}^{\mathcal{C}\cup\mathcal{P}}\right)\mid p(w\mid\boldsymbol{X}^{\mathcal{C}\cup\mathcal{P}};\bm{\beta})\right\} _{b\times1}\otimes\begin{Bmatrix}p_{\bm{\beta}_{2}}(w|\bm{x};\bm{\beta})/p(w\mid\boldsymbol{X}^{\mathcal{C}\cup\mathcal{P}};\bm{\beta})\\
\vdots\\
p_{\bm{\beta}_{w}}(w|\bm{x};\bm{\beta})/p(w\mid\boldsymbol{X}^{\mathcal{C}\cup\mathcal{P}};\bm{\beta})\\
\vdots\\
p_{\bm{\beta}_{T}}(w|\bm{x};\bm{\beta})/p(w\mid\boldsymbol{X}^{\mathcal{C}\cup\mathcal{P}};\bm{\beta})
\end{Bmatrix}_{(T-1)\times1}\right]\end{aligned}
.
\]

For example, if the optimal GPS follows a multinomial logit model, then $\bm{c}\left(w\right)$ takes the form:
\[
\begin{aligned}\bm{c}\left(w\right)= & \mathbb{E}\left[\operatorname{cov}\left\{ \boldsymbol{X}^{\mathcal{C}\cup\mathcal{P}},\mu\left(w,\boldsymbol{X}^{\mathcal{C}\cup\mathcal{P}}\right)\mid p(w\mid\boldsymbol{X}^{\mathcal{C}\cup\mathcal{P}};\bm{\beta})\right\} _{b\times1}\otimes\begin{Bmatrix}-p(2\mid\boldsymbol{X}^{\mathcal{C}\cup\mathcal{P}};\bm{\beta})\\
\vdots\\
1-p(w\mid\boldsymbol{X}^{\mathcal{C}\cup\mathcal{P}};\bm{\beta})\\
\vdots\\
-p(T\mid\boldsymbol{X}^{\mathcal{C}\cup\mathcal{P}};\bm{\beta})
\end{Bmatrix}_{(T-1)\times1}\right]\end{aligned}
.
\]

Let $p(w\mid\bm{X},\bm{\beta})$ be the GPS specification chosen by
the outcome-adjusted balance measure. Let $p(w\mid\bm{X},\hat{\bm{\beta}})$
denote the corresponding estimated GPS. We break down the variance
estimation into two parts.

First, we consider estimation of the asymptotic variance $\sigma^{2}(w)$
corresponding to matching on the true generalized propensity scores
$p(w\mid\bm{X},\bm{\beta})$. An estimator of $\sigma^{2}(w)$ is
given by
\[
\hat{\sigma}^{2}(w)=\frac{1}{N}\sum_{i=1}^{N}\left\{ \hat{Y}_{i}(w)-\hat{\mu}\left(w\right)\right\} ^{2}+\frac{1}{N}\sum_{i=1}^{N}\left[D_{i}(w)\left\{ 1+K(i,w)\right\} K(i,w)\right]\hat{\bar{\sigma}}^{2}\left\{ W_{i},p(w\mid\bm{X},\hat{\bm{\beta}})\right\} ,
\]
where
\[
\hat{\bar{\sigma}}^{2}\left\{ W_{i},p(w\mid\bm{X},\hat{\bm{\beta}})\right\} =\frac{L}{L+1}\left\{ Y_{i}-\frac{1}{L}\sum_{j\in N(i,w)}Y_{j}\right\} ^{2}.
\]
Here, $N(i,w)$ are $L\geq1$ units whose estimated generalized propensity
scores are closest to unit $i$ with the same level of treatment $w$.
Typically, $L$ is set to equal $1$.

Next, we estimate the adjustment term $\bm{c}\left(w\right)^{\text{T}}I_{\bm{\beta}}^{-1}\bm{c}\left(w\right)$.
The vector $\bm{c}(w)$ can be estimated by

\[\hat{\bm{c}}(w)=\frac{1}{N}\sum_{i=1}^{N}\widehat{\operatorname{cov}}\left\{ \bm{X}_{i},\mu(w,\bm{X}_{i})\mid p(w\mid\bm{X}_{i},\bm{\beta})\right\} \varotimes\begin{Bmatrix}p_{\bm{\beta}_{2}}(w|\bm{x};\hat{\bm{\beta}})/p(w\mid\boldsymbol{X}^{\mathcal{C}\cup\mathcal{P}};\hat{\bm{\beta}})\\
\vdots\\
p_{\bm{\beta}_{w}}(w|\bm{x};\hat{\bm{\beta}})/p(w\mid\boldsymbol{X}^{\mathcal{C}\cup\mathcal{P}};\hat{\bm{\beta}})\\
\vdots\\
p_{\bm{\beta}_{T}}(w|\bm{x};\hat{\bm{\beta}})/p(w\mid\boldsymbol{X}^{\mathcal{C}\cup\mathcal{P}};\hat{\bm{\beta}})
\end{Bmatrix},
\]
where each conditional covariance vector can be estimated by

\[
\widehat{\operatorname{cov}}\left\{ \bm{X}_{i},\mu(w,\bm{X}_{i})\mid p(w\mid\bm{X}_{i},\bm{\beta})\right\} =\frac{1}{L-1}\sum_{j\in N(i,w)}\left\{ \bm{X}_{j}-\frac{1}{L}\sum_{k\in N(i,w)}\bm{X}_{k}\right\} \left\{ Y_{j}-\frac{1}{L}\sum_{k\in N(i,w)}Y_{k}\right\} .
\]
Here, $N(i,w)$ has the same definition as the above, but $L$ is
typically set to equal $2$ or larger. The inverse of information
matrix can be estimated by the variance-covariance matrix of the fitted
GPS model parameters $\widehat{Var}(\hat{\bm{\beta}})$.

\section{Steps to perform model selection}
\begin{description}
\item [{Step$\ 0.$}] Standardize every covariate to have mean zero and
variance one. If $\boldsymbol{X}$ is high-dimensional, we recommend
applying regularized regression methods such as the Lasso (\citealp{tibshirani1996})
to screen out irrelevant covariates.
\item [{Step$\ 1.$}] Posit $K$ separate parametric models for $p(w\mid\boldsymbol{X})$,
denoted by $\left\{ p(w\mid\boldsymbol{X};\bm{\beta}_{w_{k}}):k=1,...,K\right\} $
with fixed unknown parameter vectors $\bm{\beta}_{w_{1}},...,\bm{\beta}_{w_{K}}$.
Obtain estimates $\widehat{\bm{\beta}}_{w_{1}},...,\widehat{\bm{\beta}}_{w_{K}}$
based on $\left[\left\{D_{i}(w),\boldsymbol{X}_{i}\right\}:i=1,\ldots,N\right]$. For each
unit $i$, estimate the generalized propensity scores by calculating
$\{p(w\mid\boldsymbol{X}_{i};\widehat{\bm{\beta}}_{w_{k}}):i=1,\ldots,N;k=1,...,K\}$.
\item [{Step$\ 2.$}] Choose a metric of correlation $\rho$ following the discussion above. At each treatment level $w$, compute the outcome-adjusted balance
measure for all $K$ posited models: $$\left[ \mathtt{OABM}_{\rho}\left\{w,p(w|\boldsymbol{X};\widehat{\bm{\beta}}_{w_{k}})\right\}:k=1,...,K\right].$$ 
Select the GPS model that minimizes the outcome-adjusted
balance measure for each $w$.
\item [{Step$\ 3.$}] At each treatment level $w$, obtain an estimate
of $\mathbb{E}\left\{ Y(w)\right\} $ by matching on the estimated
GPS based on the selected model from Step 2. Compute estimates of
the pairwise ATEs $\tau(w,w')$ using the GPSM estimator defined in
Equation (\ref{eq:gpsm}). 
\end{description}

\section{Existing Balance Measures}
\begin{itemize}
\item Absolute mean difference: 
\[
\mathtt{AMD}\left\{w,p(w\mid\boldsymbol{X})\right\}=\sum\limits _{j=1}^{d}\left|\widehat{X}^{(j)}_{{\rm gpsm}}(w)-\overline{X}^{(j)}\right|.
\]
\item Kolmogorov-Smirnov distance: 
\[
\mathtt{KSdist}\left\{w,p(w\mid\boldsymbol{X})\right\}=\sum\limits _{j=1}^{d}\underset{x^{(j)}}{\sup}\left|\widehat{F}_{N}(w,x^{(j)})-F_{N}(x^{(j)})\right|,
\]
where $\widehat{F}_{N}(w,x)=N^{-1}\sum\limits _{i=1}^{N}I_{\left[-\infty,x\right]}\left\{\widehat{X}_{i}(w)\right\}$ and $F_{N}(x^{(j)})=N^{-1}\sum\limits _{i=1}^{N}I_{\left[-\infty,x\right]}\left(X_{i}\right)$ are the empirical CDFs of the imputed covariate and original sample covariate respectively.
\item Mahalanobis distance is a better measure of overall balance when covariates are correlated: 
\[
\mathtt{Mdist}\left\{w,p(w\mid\boldsymbol{X})\right\}=\left[\widehat{\boldsymbol{X}}_{{\rm gpsm}}(w)-\overline{\boldsymbol{X}}\right]^{\text{T}}S^{-1}\left[\widehat{\boldsymbol{X}}_{{\rm gpsm}}(w)-\overline{\boldsymbol{X}}\right],
\]
where $\widehat{\boldsymbol{X}}_{{\rm gpsm}}(w)=\left[\widehat{X}^{(1)}_{{\rm gpsm}}(w),...,\widehat{X}^{(d)}_{{\rm gpsm}}(w)\right]^{\text{T}}$, $\overline{\boldsymbol{X}}=\left(\overline{X}^{(1)},...,\overline{X}^{(d)}\right)^{\text{T}}$, and $S$ denotes the sample covariance matrix of $\left(X^{(1)},...,X^{(d)}\right)^{\text{T}}$,
\item Weighted balance measure: 
\[
\mathtt{WBM}\left\{w,p(w\mid\boldsymbol{X})\right\}=\sum\limits _{j=1}^{d}\left|\widehat{\theta}^{(j)}\left[\widehat{X}^{(j)}_{{\rm gpsm}}(w)-\overline{X}^{(j)}\right]\right|,
\]
where $\widehat{\theta}^{(j)}$ is the coefficient of $X^{(j)}$ from a multivariate maximum likelihood regression of the observed outcome $Y$ on $\boldsymbol{X}$.
\end{itemize}

\section{Remaining Scenarios of the Simulation Study}

For the simulation study, we compare the performance of balance measures
under the combinations of (strong/weak) instrumental variables, (strong/weak)
precision variables, and potential outcome model that is a (linear/nonlinear)
function of the covariates. In this section, we focus on the remaining
scenarios where IV's are weak predictors of the treatment assignment.

\begin{figure}[p!]
\begin{centering}
\subfloat[\label{figure1a-linear-u1v1}$u=1,v=1$\protect \\
Outcome \textit{linear} in covariates]{\begin{centering}
\includegraphics[width=7cm,height=7cm]{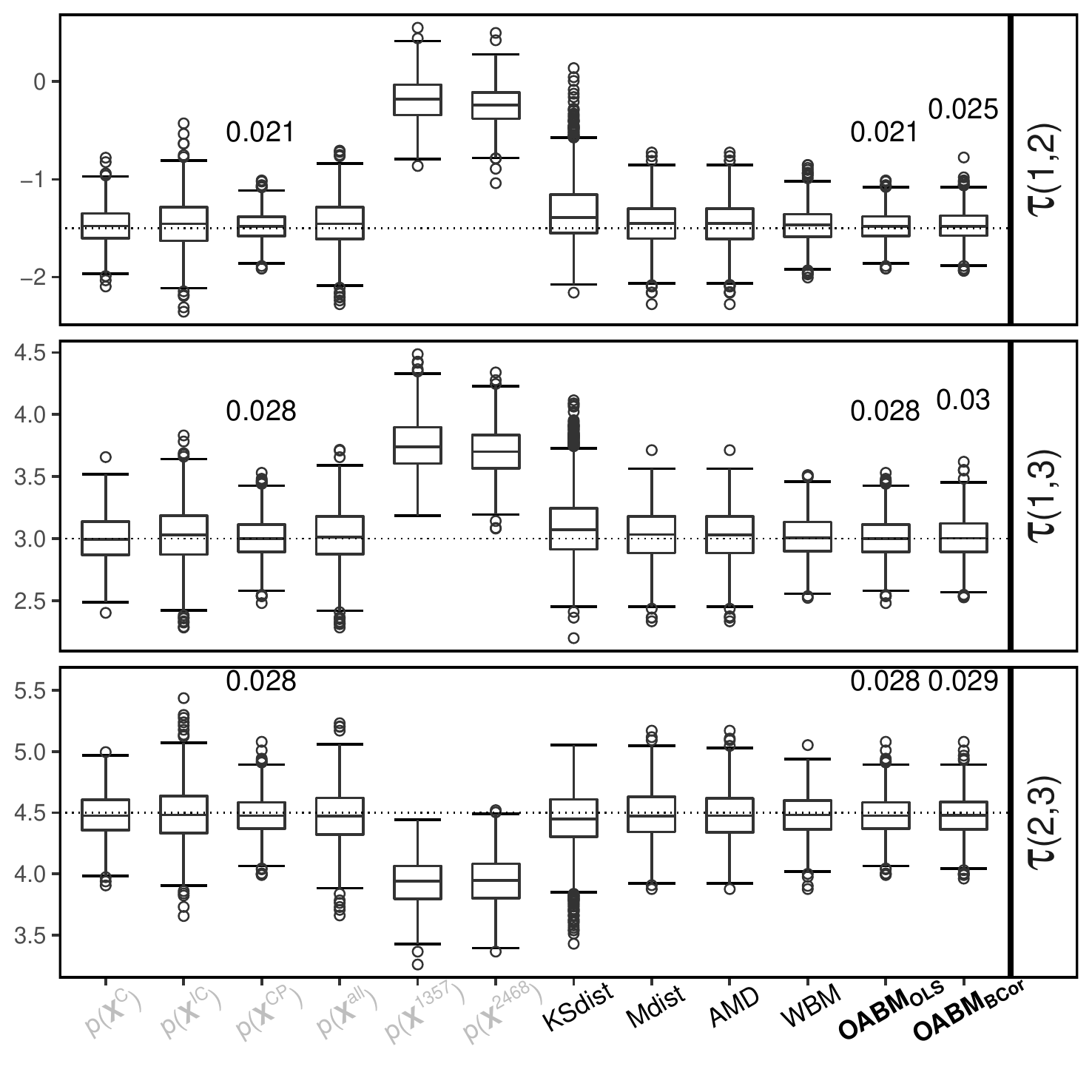}
\par\end{centering}
}\subfloat[\label{figure1b-linear-u1v2}$u=1,v=2$\protect \\
Outcome \textit{linear} in covariates]{\begin{centering}
\includegraphics[width=7cm,height=7cm]{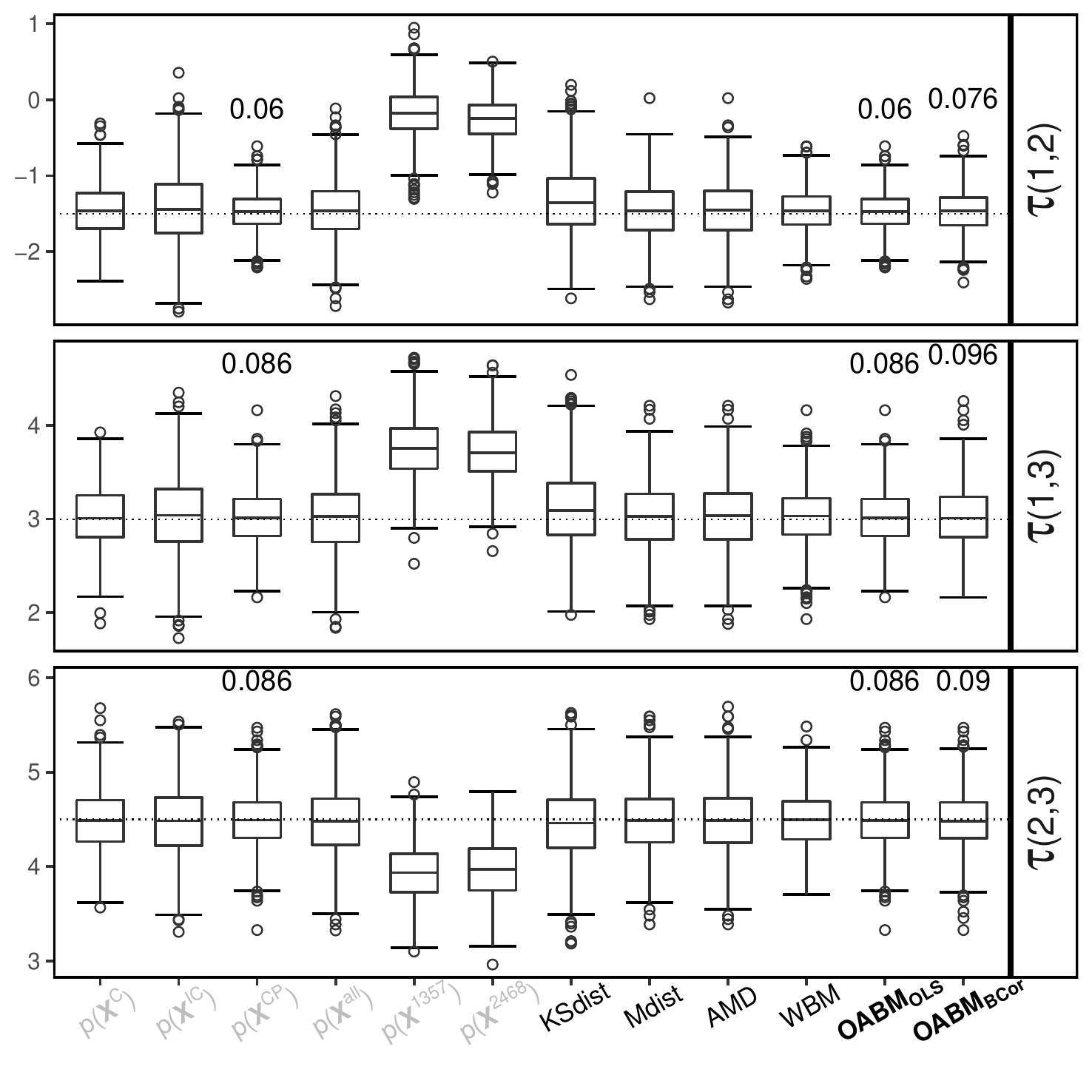}
\par\end{centering}
}
\par\end{centering}
\begin{centering}
\subfloat[\label{figure1c-nonlinear-u1v1}$u=1,v=1$\protect \\
Outcome \textit{nonlinear} in covariates]{\begin{centering}
\includegraphics[width=7cm,height=7cm]{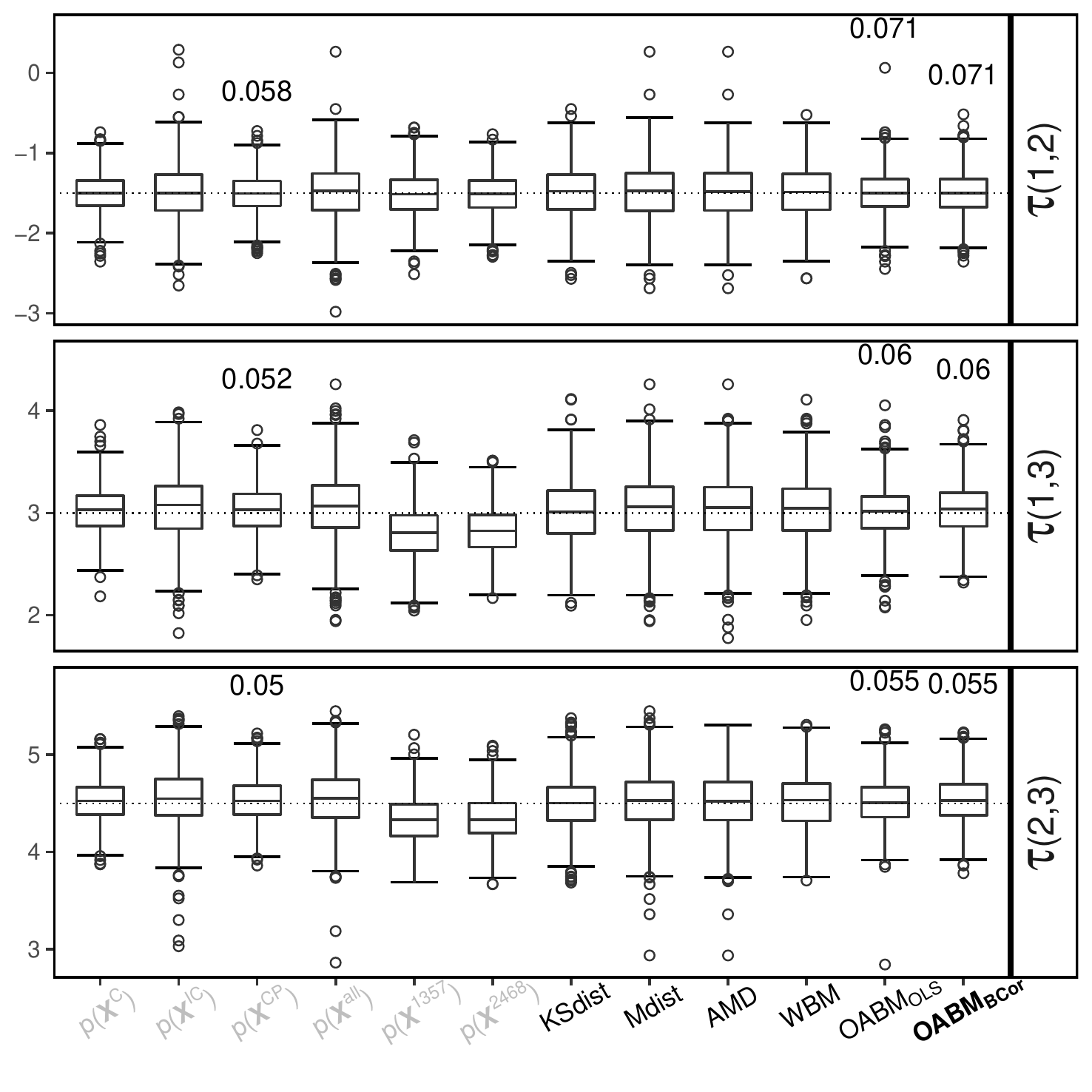}
\par\end{centering}
}\subfloat[\label{figure1d-nonlinear-u1v2}$u=1,v=2$\protect \\
Outcome \textit{nonlinear} in covariates]{\begin{centering}
\includegraphics[width=7cm,height=7cm]{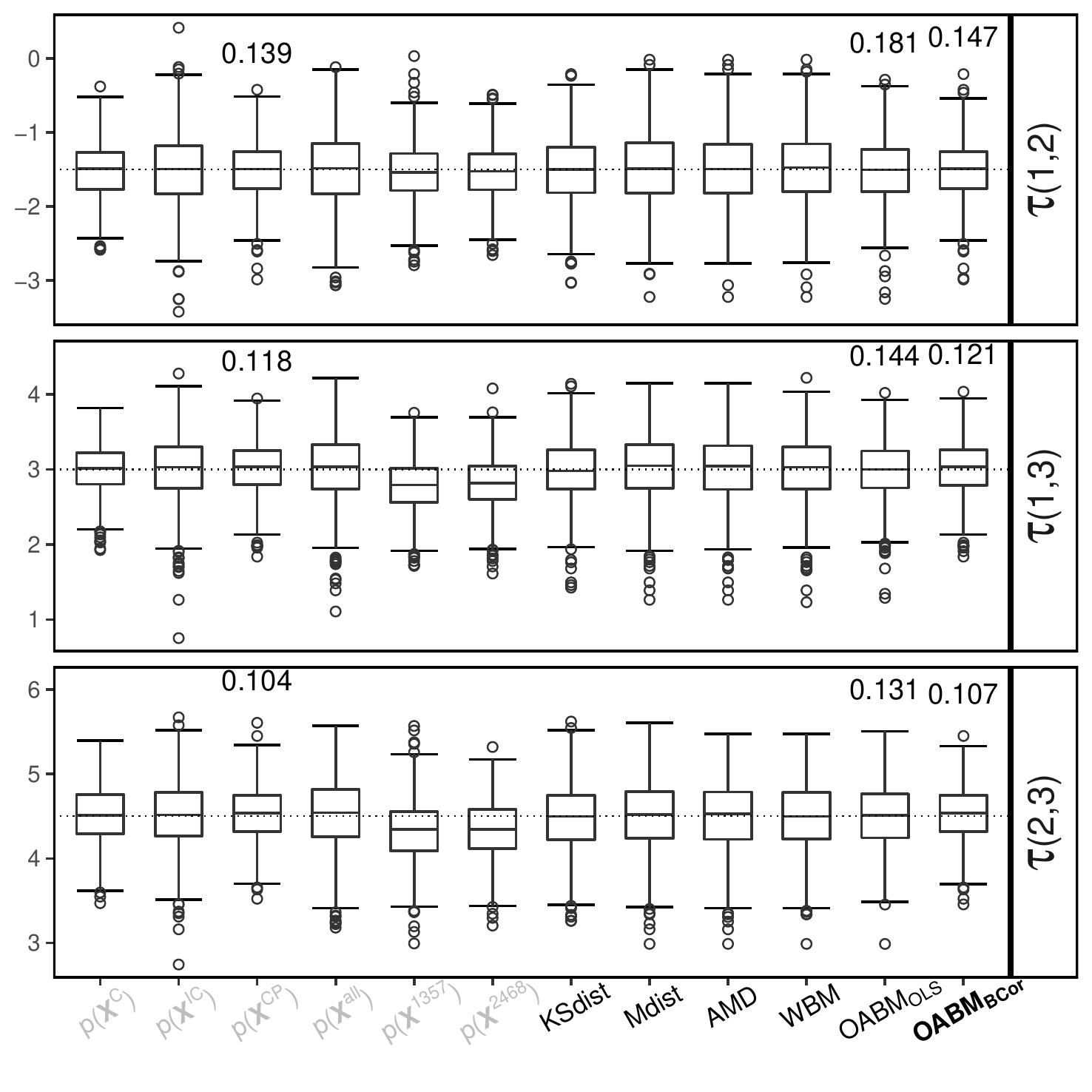}
\par\end{centering}
}
\par\end{centering}
\caption{\label{figure1-mse}Box plots of 1000 generalized propensity score
matching estimates under scenarios with weak IVs. The 6 benchmark
GPS models are greyed out on the $x$-axis. Numeric MSEs for $\mathtt{p(\bm{X}^{CP})}$,
$\mathtt{OABM_{OLS}}$, and $\mathtt{OABM_{BCor}}$ are explicitly shown above their corresponding
box plots. True pairwise treatment effects are denoted by the horizontal
dotted lines.}
\end{figure}

\begin{figure}[p!]

\begin{centering}
\subfloat[\label{figure2a-linear-u1v1}$u=1,v=1$\protect \\
Outcome \textit{linear} in covariates]{\begin{centering}
\includegraphics[width=7cm,height=7cm]{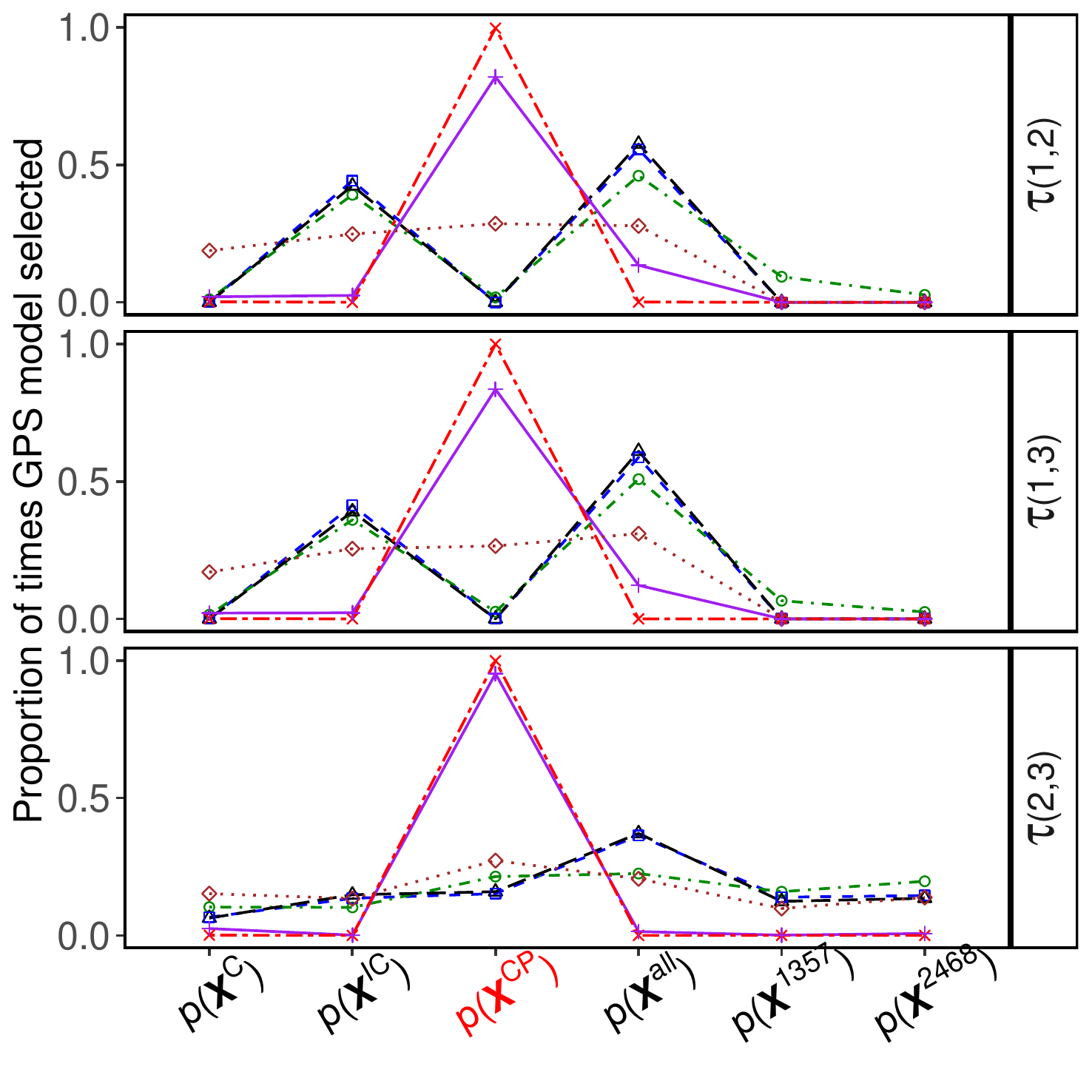}
\par\end{centering}
}\subfloat[\label{figure2b-linear-u1v2}$u=1,v=2$\protect \\
Outcome \textit{linear} in covariates]{\begin{centering}
\includegraphics[width=7cm,height=7cm]{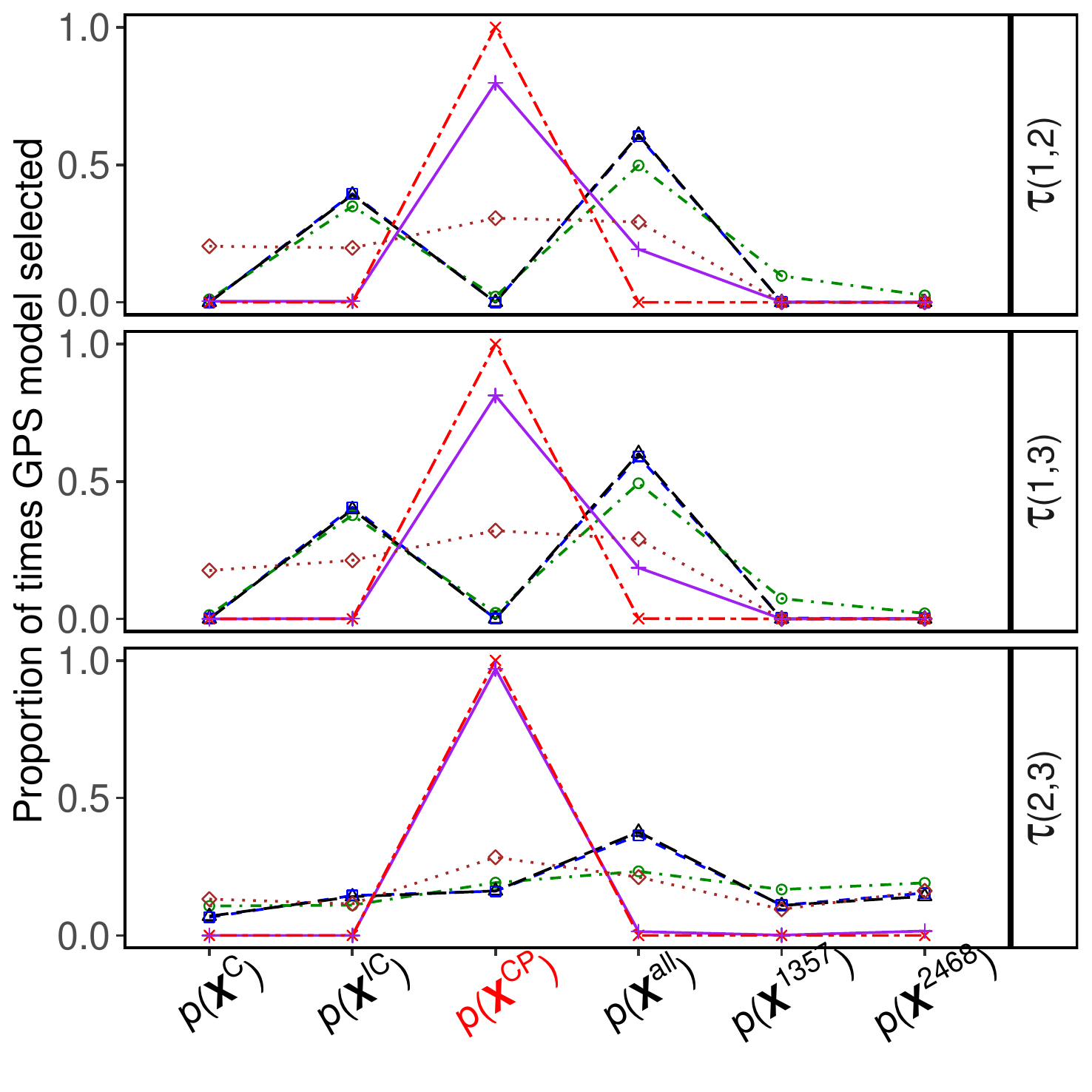}
\par\end{centering}
}
\par\end{centering}
\begin{centering}
\subfloat[\label{figure2c-nonlinear-u2v1}$u=1,v=1$\protect \\
Outcome \textit{nonlinear} in covariates]{\begin{centering}
\includegraphics[width=7cm,height=7cm]{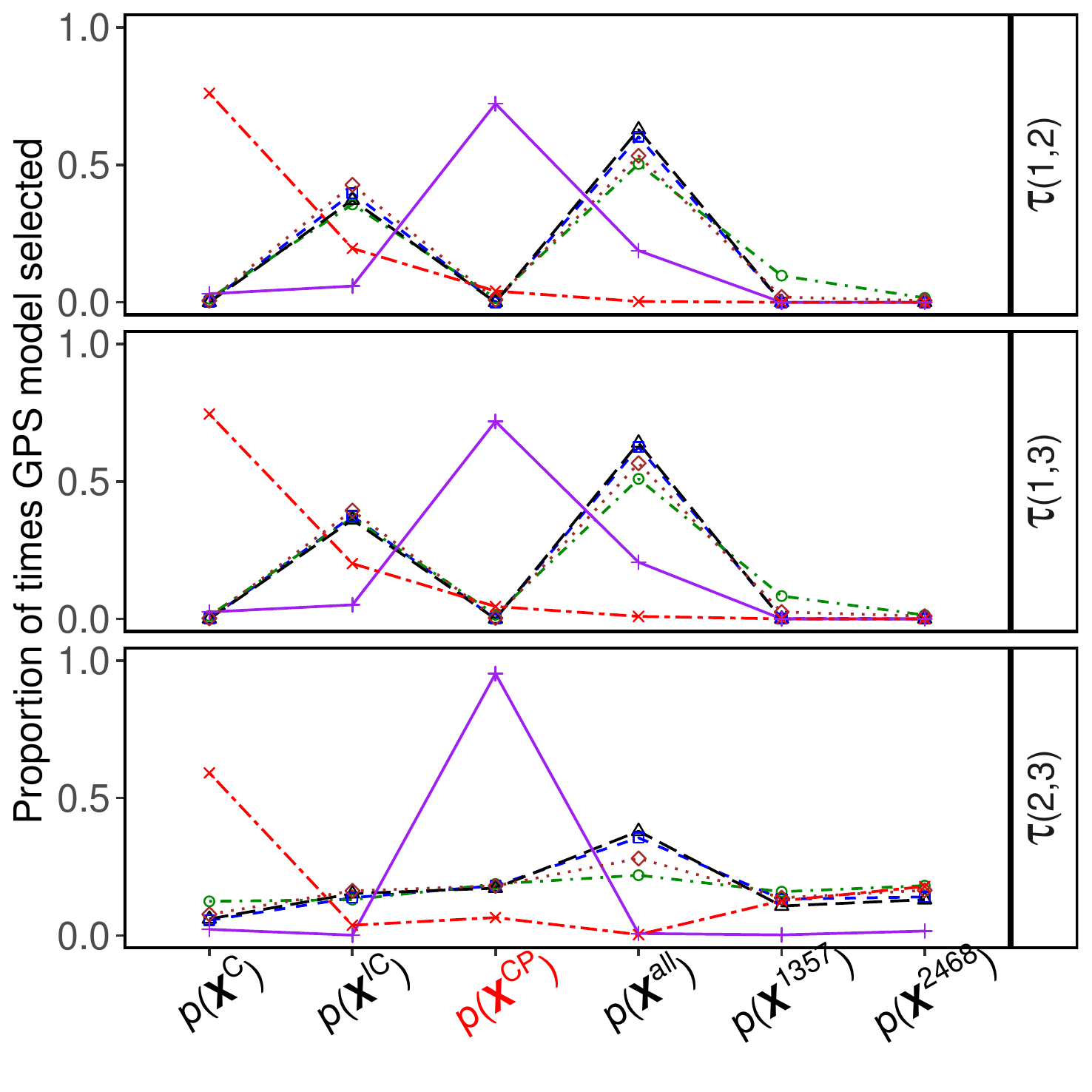}
\par\end{centering}
}\subfloat[\label{figure2d-nonlinear-u2v2}$u=1,v=2$\protect \\
Outcome \textit{nonlinear} in covariates]{\begin{centering}
\includegraphics[width=7cm,height=7cm]{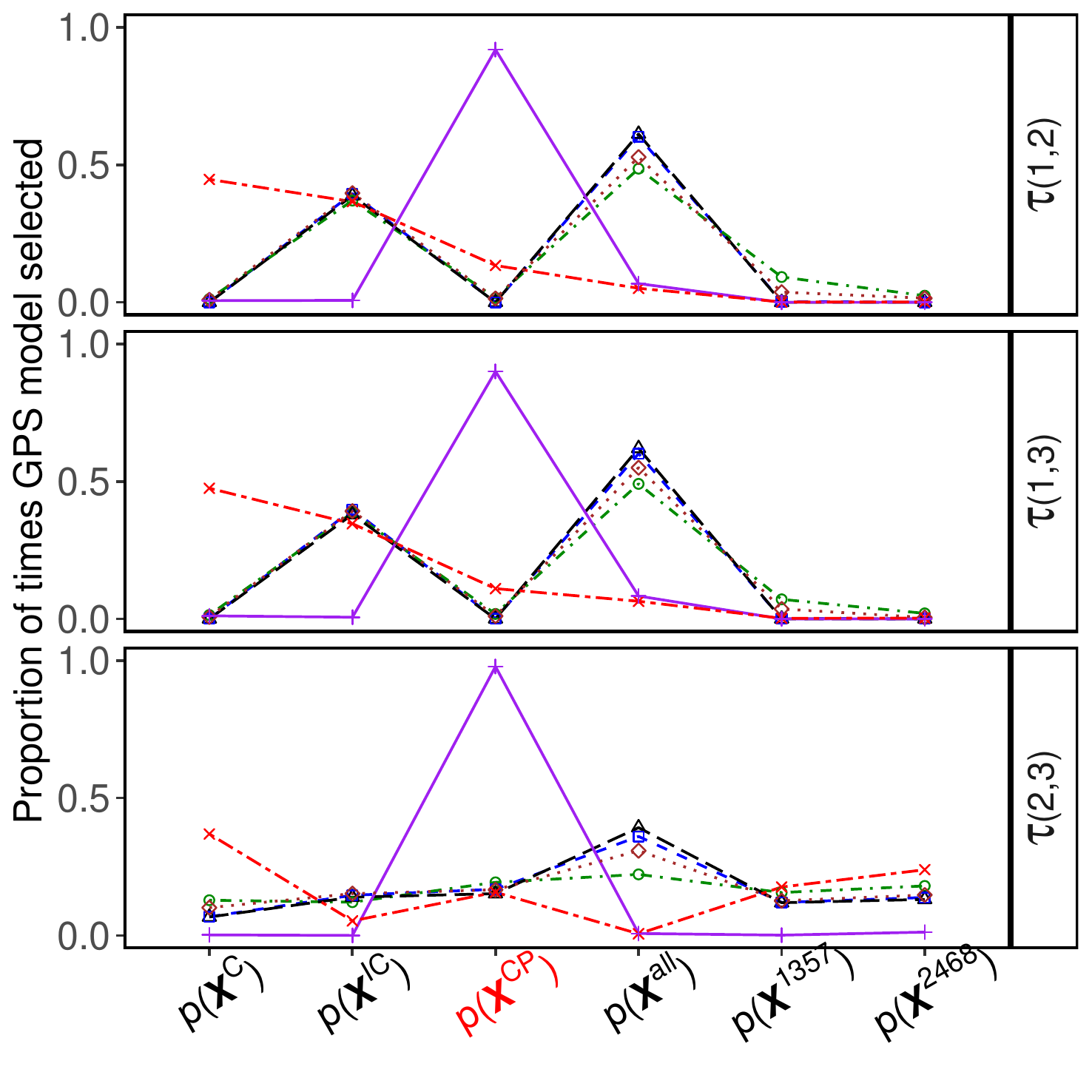}
\par\end{centering}
}
\par\end{centering}
\subfloat{\includegraphics[width=14cm,height=1.5cm]{Figures/ms_legend.pdf}}
\caption{\label{figure2-modelselection}Proportion of model selection over
1000 simulations under scenarios with weak IVs. The optimal benchmark
model $\mathtt{p(\bm{X}^{CP})}$ is colored in red.}
\end{figure}

In Figure \ref{figure1-mse}, we summarize the estimation performance
of the GPSM estimator with GPS selected based on different balance
measures. Figure \ref{figure2-modelselection} shows the percentage
of the times (out of 1000 datasets) each benchmark model is selected
by the balance measures. Table \ref{table1-coveragerates} presents
the coverage rates of the asymptotic 95\% confidence intervals.

As indicated by Figure \ref{figure1a-linear-u1v1} and \ref{figure1b-linear-u1v2},
when the covariates are linear predictors of the potential outcomes,
among all six balance measures, estimation with $\mathtt{OABM_{OLS}}$
has mean squared errors almost matching that of the optimal benchmark
model $\mathtt{p(\bm{X}^{CP})}$. This is accompanied by the fact that $\mathtt{OABM_{OLS}}$
overwhelmingly selects the optimal benchmark model, as demonstrated
by the purple line the Figure \ref{figure2a-linear-u1v1} and \ref{figure2b-linear-u1v2}.
Since the penalizing weights of $\mathtt{OABM_{OLS}}$ are computed
from a correctly specified outcome model, this result is expected
and is consistent with Theorem \ref{theorem1-model selection consistency}.
The coverage rates for all balance measures are within acceptable
range except for $\mathtt{KSdist}$, which fall short of the nominal
95\% rate by a sizable amount. Comparing to the strong IV scenarios
from the main article, the advantage of using $\mathtt{OABM_{OLS}}$
is slightly smaller when IVs are weaker, in which case including these
IVs in the GPS will result in less efficiency loss.

When the potential outcome has a nonlinear relationship with the covariates,
$\mathtt{OABM_{BCor}}$ results in slightly superior estimation performance
in terms of mean squared error, as illustrated in Figure \ref{figure2c-nonlinear-u2v1}
and \ref{figure2d-nonlinear-u2v2}. Combining the model selection
results shown in Figure 3 and \ref{figure2-modelselection},
we notice that $\mathtt{OABM_{BCor}}$ is slightly less consistent in
selecting the optimal model than $\mathtt{OABM_{OLS}}$ in the linear
cases, possibly due to the existence of collider bias. Regarding the
coverage rates, $\mathtt{OABM_{BCor}}$ also yields slightly greater
coverage than non-penalizing based balance measures, coming close
to the nominal 95\%.

\bibliographystyle{apalike}
\bibliography{refs}

\begin{table}[p!]
\captionsetup[subfloat]{justification=centering}
\centering
\subfloat[\label{table1a-linear-u1v1}$u=1,v=1$\protect \\
Outcome \textit{linear} in covariates]{
\begin{tabular}{lccc}
\hline 
 & $\tau(1,2)$ & $\tau(1,3)$ & $\tau(2,3)$\tabularnewline
\hline 
$\mathtt{KSdist}$ & 0.805 & 0.843 & 0.885\tabularnewline
$\mathtt{Mdist}$ & 0.959 & 0.950 & 0.957\tabularnewline
$\mathtt{AMD}$ & 0.956 & 0.951 & 0.956\tabularnewline
$\mathtt{WBM}$ & 0.980 & 0.977 & 0.978\tabularnewline
$\mathtt{OABM_{OLS}}$ & 0.968 & 0.950 & 0.942\tabularnewline
$\mathtt{OABM_{BCor}}$ & 0.970 & 0.955 & 0.941\tabularnewline
\hline 
\end{tabular}
}
\subfloat[\label{table1b-linear-u1v2}$u=1,v=2$\protect \\
Outcome \textit{linear} in covariates]{
\begin{tabular}{lccc}
\hline 
 & $\tau(1,2)$ & $\tau(1,3)$ & $\tau(2,3)$\tabularnewline
\hline 
$\mathtt{KSdist}$ & 0.861 & 0.857 & 0.910\tabularnewline
$\mathtt{Mdist}$ & 0.963 & 0.931 & 0.945\tabularnewline
$\mathtt{AMD}$ & 0.960 & 0.926 & 0.936\tabularnewline
$\mathtt{WBM}$ & 0.985 & 0.952 & 0.955\tabularnewline
$\mathtt{OABM_{OLS}}$ & 0.974 & 0.920 & 0.913\tabularnewline
$\mathtt{OABM_{BCor}}$ & 0.968 & 0.916 & 0.923\tabularnewline
\hline 
\end{tabular}
}

\subfloat[\label{table1c-nonlinear-u1v1}$u=1,v=1$\protect \\
Outcome \textit{nonlinear} in covariates]{
\begin{tabular}{lccc}
\hline 
 & $\tau(1,2)$ & $\tau(1,3)$ & $\tau(2,3)$\tabularnewline
\hline 
$\mathtt{KSdist}$ & 0.928 & 0.919 & 0.941\tabularnewline
$\mathtt{Mdist}$ & 0.925 & 0.909 & 0.922\tabularnewline
$\mathtt{AMD}$ & 0.923 & 0.903 & 0.925\tabularnewline
$\mathtt{WBM}$ & 0.925 & 0.904 & 0.933\tabularnewline
$\mathtt{OABM_{OLS}}$ & 0.942 & 0.941 & 0.948\tabularnewline
$\mathtt{OABM_{BCor}}$ & 0.936 & 0.934 & 0.939\tabularnewline
\hline 
\end{tabular}
}
\subfloat[\label{table1d-nonlinear-u1v2}$u=1,v=2$\protect \\
Outcome \textit{nonlinear} in covariates]{
\begin{tabular}{lccc}
\hline 
 & $\tau(1,2)$ & $\tau(1,3)$ & $\tau(2,3)$\tabularnewline
\hline 
$\mathtt{KSdist}$ & 0.945 & 0.942 & 0.941\tabularnewline
$\mathtt{Mdist}$ & 0.944 & 0.929 & 0.927\tabularnewline
$\mathtt{AMD}$ & 0.946 & 0.933 & 0.933\tabularnewline
$\mathtt{WBM}$ & 0.940 & 0.937 & 0.937\tabularnewline
$\mathtt{OABM_{OLS}}$ & 0.938 & 0.951 & 0.951\tabularnewline
$\mathtt{OABM_{BCor}}$ & 0.941 & 0.947 & 0.956\tabularnewline
\hline 
\end{tabular}
}
\caption{\label{table1-coveragerates}Coverage rates of asymptotic 95\% confidence
intervals under scenarios with weak IVs}
\end{table}

\makeatletter\@input{reference.tex}\makeatother